\documentclass[10pt]{article}
\usepackage{amsfonts,amsmath,amssymb, cite}

\usepackage{graphicx}
\usepackage{dcolumn}
\usepackage{bm}

\newcommand{\Real}{\mathop{\textrm{Re}}}
\newcommand{\Imag}{\mathop{\textrm{Im}}}

\begin{document}
\title{Gordon decomposition of the magnetizability of a Dirac one-electron atom in an arbitrary discrete energy state}
\author{Patrycja Stefa{\'n}ska \\*[3ex]
Institute of Physics and Applied Computer Science, \\
Faculty of Applied Physics and Mathematics,
Gda{\'n}sk University of Technology, \\
Narutowicza 11/12, 80--233 Gda{\'n}sk, Poland \\
Email: patrycja.stefanska@pg.edu.pl}
\date{\today}
\maketitle
\begin{center}
\textbf{Published as: Atoms 12 (2024) 54}
\\*[1ex]
\textbf{doi: 10.3390/atoms12110054} \\*[5ex]
\end{center}
\begin{abstract}
We present a Gordon decomposition of the magnetizability of a Dirac one-electron atom in an arbitrary discrete energy eigenstate, with a pointlike, spinless, and motionless nucleus of charge $Ze$. The external magnetic field, by which the atomic state is perturbed, is assumed to be weak, static, and uniform. Using the Sturmian expansion of the generalized Dirac--Coulomb Green function proposed by Szmytkowski in 1997, we derive a closed-form expressions for the diamagnetic ($\chi_{d}$) and paramagnetic ($\chi_{p}$) contributions to $\chi$. Our calculations are purely analytical; the received formula for $\chi_{p}$ contains the generalized hypergeometric functions ${}_3F_2$ of the unit argument, while $\chi_{d}$ is of an elementary form. For the atomic ground state, both results reduce to the formulas obtained earlier by other author. This work is a prequel to our recent article, where the numerical values of $\chi_{d}$ and $\chi_{p}$ for some excited states of selected hydrogenlike ions with $1 \leqslant Z \leqslant 137$ were obtained with the use of the general formulas derived here.

\vskip3ex


\end{abstract}
%
%
\section{Introduction}
\label{I}
\setcounter{equation}{0}

The influence of external electric and magnetic fields on atoms is one of the most interesting physical phenomena widely discussed in the literature. In the theoretical approach, it can be described by changing two physical quantities, namely the electric charge density and the electric current density. These changes consist mainly of inducing additional multipole moments in the system which, in the first approximation, are proportional to the intensity of the applied field; the proportionality factors can be expressed by means of appropriate electromagnetic susceptibilities \cite{Saku67, Gord28}. In nonrelativistic quantum mechanics, many second-order {\emph{magnetic}} roperties are obtained as a sum of a {\emph{diamagnetic}} and a {\emph{paramagnetic}} term \cite{Vleck32,Hodg14,Summ24}. By performing the decomposition into these terms, one can often indicate that the paramagnetic (or Van Vleck) term vanishes, or is very small in magnitude. Then, the dominating diamagnetic contribution can easily be evaluated in terms of the unperturbed wave function only. 

Finding the paramagnetic and (especially) diamagnetic contributions in the framework of relativistic theory is not unequivocal, in the sense that many different techniques may be utilized to achieve this. For example, some authors used a separate summation over negative-energy states \cite{Ster62, Pyyk77, Pyyk83, Auca99}, unitary transformation of the Dirac operator \cite{Kutz03}, or Gordon decomposition of the electric current (or charge) density \cite{Szmy02, Pype83, Pype83a, Pype88, Pype99, Pype99a, Batt01}. In our opinion, the latter method is the most elegant,  and has the potential to be applied to many physical issues.

In most of the articles, the Gordon decomposition was employed to considerations of the susceptibilities of hydrogenlike atoms, and concerned the atomic {\emph{ground state}} only. 
In this work, we shall use this approach to find the diamagnetic and paramagnetic contributions to the magnetizability of a relativistic one-electron atom being in an {\emph{arbitrary}} discrete energy eigenstate. In calculations of such a general formula, we have used perturbation theory combined with the Green function technique. To derive the explicit expressions for dia- and paramagnetic terms, we have utilized expansion of the first-order generalized Dirac--Coulomb Green function (DCGF) in the Sturmian basis \cite{Szmy97}; this form of Green's function was widely used by us in various calculations of the electromagnetic properties of a Dirac one-electron atom \cite{Stef15, Stef16, Stef16a, Stef16b, Stef17, Stef18} (for other applications of this representation of DCGF, see refs. \cite{Szmy02b, Szmy04, Szmy11, Stef12, Szmy12, Szmy14}).  For the atom in the ground state, our general results presented in this work reduce to the formulas obtained earlier by Szmytkowski, in ref. \cite{Szmy02b}.

\section{Preliminaries}
\label{II}
\setcounter{equation}{0}
The system we shall be concerned with in the present work is a Dirac one-electron atom with a spinless, pointlike, and infinitely heavy nucleus of charge $+Ze$, and with an electron of mass $m_e$ and charge $-e$. Without the external perturbations, the atomic state energy {levels are} 
\begin{equation}
E^{(0)} \equiv E_{n \kappa}^{(0)}=m_ec^2\frac{n+\gamma_{\kappa}}{N_{n \kappa}},
\label{2.1}
\end{equation}
with
\begin{equation}
N_{n\kappa}=\sqrt{n^2+2n\gamma_{\kappa}+\kappa^2}
\label{2.2}
\end{equation}
and
\begin{equation}
\gamma_{\kappa}=\sqrt{\kappa^2-(\alpha Z)^2},
\label{2.3}
\end{equation}
where $n$ is the radial quantum number,  $\kappa=\pm1, \pm2, \ldots$ is the Dirac quantum number, and $\alpha$ denotes the Sommerfeld's fine-structure constant. The atomic energy levels from Equation~(\ref{2.1}) are degenerate; the normalized to unity eigenfunctions associated with the eigenvalue $E_{n \kappa}^{(0)}$ are given by
\begin{equation}
\Psi^{(0)}(\boldsymbol{r}) \equiv \Psi_{n \kappa \mu}^{(0)}(\boldsymbol{r}) 
=\frac{1}{r} 
\left(
\begin{array} {c}
P_{n\kappa}^{(0)}(r) \Omega_{\kappa\mu}(\boldsymbol{n}_r) \\ 
\textrm{i} Q_{n\kappa}^{(0)}(r) \Omega_{-\kappa\mu}(\boldsymbol{n}_r) 
\end{array} 
\right),
\label{2.4}
\end{equation}
where $\Omega_{\kappa \mu}(\boldsymbol{n}_r)$ (with $\boldsymbol{n}_r=\boldsymbol{r}/r$ and $\mu=-|\kappa|+\frac{1}{2}, -|\kappa|+\frac{3}{2}, \ldots , |\kappa|-\frac{1}{2}$) are the orthonormal spherical spinors defined as in ref. \cite{Szmy07}, while the radial functions are normalized to unity, in the sense of
\begin{equation}
\int_0^{\infty} \textrm{d}r 
\left\{
[P_{n\kappa}^{(0)}(r)]^2+[Q_{n\kappa}^{(0)}(r)]^2 
\right\}
=1
\label{2.5}
\end{equation}
and are explicitly given by
\begin{eqnarray}
P_{n\kappa}^{(0)}(r)=\sqrt{\frac{Z}{2a_0} \frac{(1+\epsilon_{n \kappa}) (n+2\gamma_{\kappa}) n!}{N_{n\kappa}^2(N_{n\kappa}-\kappa)\Gamma(n+2\gamma_{\kappa})}}  
\left(
\frac{2Zr}{a_{0}N_{n\kappa}}\right)^{\gamma_{\kappa}}\exp\left(\frac{-Zr}{a_0N_{n\kappa}}\right)
\nonumber \\
\times 
\left[
L_{n-1}^{(2\gamma_{\kappa})}\left(\frac{2Zr}{a_{0}N_{n\kappa}}\right)
+\frac{\kappa-N_{n\kappa}}{n+2\gamma_{\kappa}}L_{n}^{(2\gamma_{\kappa})}\left(\frac{2Zr}{a_{0}N_{n\kappa}}\right)
\right],
\label{2.6}
\end{eqnarray}
\begin{eqnarray}
Q_{n\kappa}^{(0)}(r)=\sqrt{\frac{Z}{2a_0}\frac{ (1-\epsilon_{n \kappa}) (n+2\gamma_{\kappa}) n!}{N_{n\kappa}^2(N_{n\kappa}-\kappa)\Gamma(n+2\gamma_{\kappa})}} 
\left(\frac{2Zr}{a_{0}N_{n\kappa}}\right)^{\gamma_{\kappa}}  \exp\left(\frac{-Zr}{a_0N_{n\kappa}}\right)
\nonumber \\
\times 
\left[
L_{n-1}^{(2\gamma_{\kappa})} \left(\frac{2Zr}{a_{0}N_{n\kappa}}\right) 
-\frac{\kappa-N_{n\kappa}}{n+2\gamma_{\kappa}}L_{n}^{(2\gamma_{\kappa})}\left(\frac{2Zr}{a_{0}N_{n\kappa}}\right)
\right]. 
\label{2.7}
\end{eqnarray}
Here $L_{n}^{(\beta)}(\rho)$ denote the generalized Laguerre polynomial \cite{Magn66}, $a_0$ is the Bohr radius, and 
\begin{equation}
\epsilon_{n \kappa}=\frac{E_{n\kappa}^{(0)}}{m_ec^2}=\frac{n+\gamma_{\kappa}}{N_{n\kappa}}.
\label{2.8}
\end{equation}

Suppose that the atom is placed in  a weak, static, uniform magnetic field of induction $\boldsymbol{B}$ directed along the $z$ axis of a Cartesian coordinate system. The time-independent Dirac equation for the atomic electron is then
\begin{equation}
\left[
-\mathrm{i}c\hbar\boldsymbol{\alpha}\cdot\boldsymbol{\nabla}
+e c \boldsymbol{\alpha}\cdot\boldsymbol{A}(\boldsymbol{r})+\beta
m_ec^{2}+V_c-E
\right]
\Psi(\boldsymbol{r})=0
\label{2.9}
\end{equation}
 supplemented by the boundary conditions
\begin{equation}
r \Psi(\boldsymbol{r}) \stackrel{r \to 0}{\longrightarrow}0, 
\qquad \qquad 
r^{3/2} \Psi(\boldsymbol{r}) \stackrel{r \to \infty}{\longrightarrow}0. 
\label{2.10}
\end{equation}
In Equation (\ref{2.9}), $\boldsymbol{\alpha}$ and $\beta$ are the standard Dirac matrices \cite{Schi55}, and $V_c=-{Ze^{2}}/{(4\pi\epsilon_{0})r}$ is the Coulomb potential, whereas the vector potential $\boldsymbol{A}(\boldsymbol{r})$, written in a symmetric gauge, has the form
\begin{equation}
\boldsymbol{A}(\boldsymbol{r})=\frac{1}{2} \boldsymbol{B}\times\boldsymbol{r}.
\label{2.11}
\end{equation}

By virtue of our assumption that the external magnetic field is weak,  the interaction~operator
\begin{equation}
\hat{H}^{(1)}=e c \boldsymbol{\alpha}\cdot\boldsymbol{A}(\boldsymbol{r})
\label{2.12}
\end{equation}
may be treated a small perturbation of the Dirac--Coulomb Hamiltonian describing an isolated atom. The corresponding zeroth-order bound state eigenproblem is given by the equation
\begin{equation}
\left[
-\mathrm{i}c\hbar\boldsymbol{\alpha}\cdot\boldsymbol{\nabla}
+\beta m_e c^{2}+V_c-E^{(0)}
\right]
\Psi^{(0)}(\boldsymbol{r})=0,
\label{2.13}
\end{equation}
with the boundary conditions
\begin{equation}
r \Psi^{(0)}(\boldsymbol{r}) \stackrel{r \to 0}{\longrightarrow}0, 
\qquad \qquad 
r^{3/2} \Psi^{(0)}(\boldsymbol{r}) \stackrel{r \to \infty}{\longrightarrow}0. 
\label{2.14}
\end{equation}
The functions from Equation (\ref{2.4}) are adjusted to the perturbation $\hat{H}^{(1)}$, i.e., they diagonalize the matrix  of that perturbation. Thus, the solutions of eigenproblem (\ref{2.9})--(\ref{2.10}), to the lowest order in the perturbing field, can be approximated as
\begin{equation}
\Psi(\boldsymbol{r}) \simeq \Psi^{(0)}(\boldsymbol{r})+\Psi^{(1)}(\boldsymbol{r})
\label{2.15}
\end{equation}
and
\begin{equation}
E \simeq E^{(0)}+E^{(1)}.
\label{2.16}
\end{equation}
We assume that the corrections $\Psi^{(1)}(\boldsymbol{r})$ and $E^{(1)}$ are small quantities of the first order in the magnetic field strength $B=|\boldsymbol{B}|$; they solve the inhomogeneous differential equation
\begin{equation}
\left[
-\textrm{i} c \hbar \boldsymbol{\alpha} \cdot \boldsymbol{\nabla} 
+\beta m_ec^2  
+V_c
-E^{(0)} 
\right]
\Psi^{(1)}(\boldsymbol{r})
=- 
\left[
\frac{1}{2} e c \boldsymbol{{B}} \cdot 
\left(
\boldsymbol{r} \times \boldsymbol{\alpha}
\right)
-E^{(1)} 
\right] 
\Psi^{(0)}(\boldsymbol{r}) 
\label{2.17}
\end{equation}
supplemented by the boundary conditions
\begin{equation}
r \Psi^{(1)}(\boldsymbol{r}) \stackrel{r \to 0}{\longrightarrow}0, 
\qquad \qquad
 r^{3/2} \Psi^{(1)}(\boldsymbol{r}) \stackrel{r \to \infty}{\longrightarrow}0. 
\label{2.18}
\end{equation}
After carrying out some transformations from the standard perturbation theory, taking advantage of the orthogonality constraint
\begin{equation}
\int_{\mathbb{R}^3} \textrm{d}^3\boldsymbol{r}\: \Psi^{(0)\dagger}(\boldsymbol{r}) \Psi^{(1)}(\boldsymbol{r})=0,
\label{2.19}
\end{equation}
we find
\begin{equation}
E^{(1)} \equiv E_{n\kappa\mu}^{(1)}=\frac{1}{2} ec \boldsymbol{B} \cdot \int_{\mathbb{R}^3} \textrm{d}^3\boldsymbol{r} \: \Psi_{n \kappa \mu}^{(0)\dagger}(\boldsymbol{r}) 
\left(
\boldsymbol{r} \times \boldsymbol{\alpha} 
\right) 
\Psi_{n \kappa \mu}^{(0)}(\boldsymbol{r})
\label{2.20}
\end{equation}
and
\begin{eqnarray}
\Psi^{(1)}(\boldsymbol{r}) \equiv \Psi_{n \kappa \mu}^{(1)}(\boldsymbol{r})= 
-\int_{\mathbb{R}^3} \textrm{d}^3\boldsymbol{r}' \: \bar{G}^{(0)}(\boldsymbol{r},\boldsymbol{r}') 
\left[ 
\frac{1}{2} e c \boldsymbol{B} \cdot 
\left( 
\boldsymbol{r}' \times \boldsymbol{\alpha}
\right)
-E_{n \kappa \mu}^{(1)} 
\right] 
\Psi_{n \kappa \mu}^{(0)}(\boldsymbol{r}').
\label{2.21}
\end{eqnarray}
Here,  $\bar{G}^{(0)}(\boldsymbol{r},\boldsymbol{r}')$ is the generalized Dirac--Coulomb Green function associated with the energy level $E_{n \kappa}^{(0)}$ of an isolated atom. It is defined as a solution of the following differential equation (for a fixed $\boldsymbol{r}'$):
\begin{equation}
\left[-\textrm{i} c \hbar \boldsymbol{\alpha} \cdot \boldsymbol{\nabla} + \beta m_ec^2 +V_c  - E_{n\kappa}^{(0)}  \right] \bar{G}^{(0)}(\boldsymbol{r},\boldsymbol{r}') 
=\mathcal{I} \delta^3(\boldsymbol{r}-\boldsymbol{r}')-\sum_{\substack{\kappa' \mu \\ (|\kappa'|=|\kappa|)}}\Psi_{n \kappa' \mu}^{(0)}(\boldsymbol{r}) \Psi_{n \kappa' \mu}^{(0)\dagger}(\boldsymbol{r}'), 
\label{2.22}
\end{equation}
with the boundary conditions
\begin{equation}
r \bar{G}^{(0)}(\boldsymbol{r},\boldsymbol{r}') \stackrel{r \to 0}{\longrightarrow}0, \qquad  \qquad r \bar{G}^{(0)}(\boldsymbol{r},\boldsymbol{r}') \stackrel{r \to \infty}{\longrightarrow}0.
\label{2.23}
\end{equation}
Since it is Hermitian in the sense of
\begin{equation}
\bar{G}^{(0)}(\boldsymbol{r},\boldsymbol{r}')=\bar{G}^{(0)\dagger}(\boldsymbol{r}',\boldsymbol{r})
\label{2.24}
\end{equation}
and satisfies the orthogonality condition,
	\begin{equation}
\int_{\mathbb{R}^3} \textrm{d}^3\boldsymbol{r}  \Psi_{n \kappa' \mu}^{(0)\dagger}(\boldsymbol{r})  \bar{G}^{(0)}(\boldsymbol{r},\boldsymbol{r}')=0 \qquad \qquad (\textrm{for} \ \ \kappa'= \pm \kappa),
\label{2.25}
\end{equation}
the Formula (\ref{2.21}) for the first order correction to the wave function becomes
\begin{equation}
\Psi_{n \kappa \mu}^{(1)}(\boldsymbol{r})=-\frac{1}{2} e c \boldsymbol{B} \cdot \int_{\mathbb{R}^3} \textrm{d}^3\boldsymbol{r}' \:\bar{G}^{(0)}(\boldsymbol{r},\boldsymbol{r}') 
\left( 
\boldsymbol{r}' \times \boldsymbol{\alpha}
\right) 
\Psi_{n \kappa \mu}^{(0)}(\boldsymbol{r}').
\label{2.26}
\end{equation}

In order to convert the Equation (\ref{2.26}) into another form (which will be more convenient for calculations provided in the following chapters), we shall rewrite Equations (\ref{2.13}) and (\ref{2.22}) as
\begin{equation}
\Psi_{n \kappa \mu}^{(0)}(\boldsymbol{r})=\frac{\textrm{i} \hbar}{m_e c} \beta \boldsymbol{\alpha} \cdot \boldsymbol{\nabla}\Psi_{n \kappa \mu}^{(0)}(\boldsymbol{r})+\frac{1}{m_e c^2}\left[E^{(0)}-V_c(r)\right] \beta \Psi_{n \kappa \mu}^{(0)}(\boldsymbol{r})
\label{2.27}
\end{equation}
and
\begin{eqnarray}
\bar{G}^{(0)}(\boldsymbol{r},\boldsymbol{r}')&=&\frac{\textrm{i} \hbar}{m_e c} \beta \boldsymbol{\alpha} \cdot \boldsymbol{\nabla} \bar{G}^{(0)}(\boldsymbol{r},\boldsymbol{r}')
+\frac{1}{m_e c^2}\left[E^{(0)}-V_c(r)\right] \beta \bar{G}^{(0)}(\boldsymbol{r},\boldsymbol{r}') 
\nonumber
\\*[1ex]
\quad
&&
+\frac{1}{m_e c^2} \beta \delta^3(\boldsymbol{r}-\boldsymbol{r}')
 -\frac{1}{m_e c^2} \sum_{\substack{\kappa' \mu \\ (|\kappa'|=|\kappa|)}}\beta \Psi_{n \kappa' \mu}^{(0)}(\boldsymbol{r}) \Psi_{n \kappa' \mu}^{(0)\dagger}(\boldsymbol{r}'), 
\label{2.28}
\end{eqnarray}
respectively. If the variables $\boldsymbol{r}$ and $\boldsymbol{r}'$ will be swapped in the Equation (\ref{2.28}) and the resulting formula will be Hermitian conjugated, then by using the relation (\ref{2.24}), we will obtain
\begin{eqnarray}
\bar{G}^{(0)}(\boldsymbol{r},\boldsymbol{r}')&=&-\frac{\textrm{i} \hbar}{m_e c} \boldsymbol{\nabla}' \bar{G}^{(0)}(\boldsymbol{r},\boldsymbol{r}')  \cdot  \boldsymbol{\alpha} \beta   +\frac{1}{m_e c^2}\left[E^{(0)}-V_c(r')\right] \bar{G}^{(0)}(\boldsymbol{r},\boldsymbol{r}') \beta
\nonumber
\\*[1ex]
\quad
&&
+\frac{1}{m_e c^2} \beta \delta^3(\boldsymbol{r}'-\boldsymbol{r})
 -\frac{1}{m_e c^2} \sum_{\substack{\kappa' \mu \\ (|\kappa'|=|\kappa|)}}\Psi_{n \kappa' \mu}^{(0)}(\boldsymbol{r}) \Psi_{n \kappa' \mu}^{(0)\dagger}(\boldsymbol{r}') \beta. 
\label{2.29}
\end{eqnarray}
Now, taking advantage of the obvious identity $1=\frac{1}{2}+\frac{1}{2}$, we may rewrite Equations (\ref{2.26}) as
\begin{eqnarray}
\Psi_{n \kappa \mu}^{(1)}(\boldsymbol{r})&=&-\frac{1}{4} e c \boldsymbol{B} \cdot \int_{\mathbb{R}^3} \textrm{d}^3\boldsymbol{r}' \:\bar{G}^{(0)}(\boldsymbol{r},\boldsymbol{r}') 
\left( 
\boldsymbol{r}' \times \boldsymbol{\alpha}
\right) 
\Psi_{n \kappa \mu}^{(0)}(\boldsymbol{r}')
\nonumber
\\*[1ex]
&& \quad -\frac{1}{4} e c \boldsymbol{B} \cdot \int_{\mathbb{R}^3} \textrm{d}^3\boldsymbol{r}' \:\bar{G}^{(0)}(\boldsymbol{r},\boldsymbol{r}') 
\left( 
\boldsymbol{r}' \times \boldsymbol{\alpha}
\right) 
\Psi_{n \kappa \mu}^{(0)}(\boldsymbol{r}').
\label{2.30}
\end{eqnarray}
Inserting Equation (\ref{2.27}) into the first component on the right-hand side of the above equation, and substituting  Equation (\ref{2.29}) into the second one, with some algebra, we obtain
\begin{eqnarray}
\Psi_{n \kappa \mu}^{(1)}(\boldsymbol{r})&=&\frac{e B}{4 m_e c} \boldsymbol{n}_{z} \cdot \left( \boldsymbol{r} \times \boldsymbol{\alpha}\right) \beta  \Psi_{n \kappa \mu}^{(0)}(\boldsymbol{r}) 
-\frac{e \hbar B}{2m_e}  \int_{\mathbb{R}^3} \textrm{d}^3\boldsymbol{r}' \:\bar{G}^{(0)}(\boldsymbol{r},\boldsymbol{r}') \boldsymbol{n}_{z} \cdot \left(\hat{\boldsymbol{\Lambda}}'+\boldsymbol{\Sigma}\right) \beta \Psi_{n \kappa \mu}^{(0)}(\boldsymbol{r}')
\nonumber
\\*[1ex]
&& \quad
-
\frac{e  B}{4 m_e c}  \Psi_{n \kappa \mu}^{(0)}(\boldsymbol{r})   \int_{\mathbb{R}^3} \textrm{d}^3\boldsymbol{r}' \:  \Psi_{n \kappa \mu}^{(0)\dagger}(\boldsymbol{r}')  \boldsymbol{n}_{z} \cdot \left( \boldsymbol{r}' \times \boldsymbol{\alpha}\right) \beta \Psi_{n \kappa \mu}^{(0)}(\boldsymbol{r}').
\label{2.31}
\end{eqnarray}
Here, $\boldsymbol{\Sigma}$ is the vector matrix
\begin{eqnarray}
\boldsymbol{\Sigma}=
\left(
\begin{array}{cc}
\boldsymbol{\sigma} & 0 \\
0 & \boldsymbol{\sigma}
\end{array}
\right),
\label{2.32}
\end{eqnarray}
with $\boldsymbol{\sigma}$ being a vector composed of Pauli matrices, while
\begin{equation}
\hat{\boldsymbol{\Lambda}}'=-\textrm{i} \boldsymbol{r}' \times \boldsymbol{\nabla}.
\label{2.33}
\end{equation}
In the rest of this paper, for readability, we shall omit all the subscripts at $\Psi_{n\kappa \mu}^{(0)}$ and $\Psi_{n\kappa \mu}^{(1)}$.

\section{Gordon decomposition of the induced current density}
\label{III}
\setcounter{equation}{0}

The process of decomposing of the atomic magnetizability into diamagnetic and paramagnetic parts, which is the main goal of the present work,  requires finding the analogous components of the induced current density first. Thus, in this section, following by ideas of Pyper \cite{Pype83, Pype83a, Pype88, Pype99, Pype99a} and Szmytkowski \cite{Szmy02}, we shall provide the Gordon decomposition of $\boldsymbol{j}^{(1)}(\boldsymbol{r})$ into the following sum:
\begin{equation}
\boldsymbol{j}^{(1)}(\boldsymbol{r})=\boldsymbol{j}_{d}^{(1)}(\boldsymbol{r})+\boldsymbol{j}_{p}^{(1)}(\boldsymbol{r}),
\label{3.1}
\end{equation}
where $\boldsymbol{j}_{d}^{(1)}(\boldsymbol{r})$ and  $\boldsymbol{j}_{p}^{(1)}(\boldsymbol{r})$ we define as the diamagnetic and paramagnetic contributions to the induced current density, respectively. 

In the case under consideration when the system is the relativistic hydrogenlike atom in the static, uniform, and weak magnetic field, characterized briefly in the preceding section, the current density is given by
\begin{equation}
\boldsymbol{j}(\boldsymbol{r})=-ec \Psi^{\dagger}(\boldsymbol{r}) \boldsymbol{\alpha} \Psi(\boldsymbol{r}),
\label{3.2}
\end{equation}
provided the wave function $\Psi(r)$ is normalized to unity:
\begin{equation}
\int_{\mathbb{R}^3} \textrm{d}^3\boldsymbol{r} \: \Psi^{\dagger}(\boldsymbol{r})  \Psi(\boldsymbol{r}) =1.
\label{3.3}
\end{equation}
Equations (\ref{3.2}) and (\ref{2.15}), together with the orthogonality constraint (\ref{2.19}), imply that to the first order in the perturbing magnetic field the current $\boldsymbol{j}(\boldsymbol{r})$ can be approximated as
\begin{equation}
\boldsymbol{j}(\boldsymbol{r}) \simeq \boldsymbol{j}^{(0)}(\boldsymbol{r})+\boldsymbol{j}^{(1)}(\boldsymbol{r}),
\label{:3.4}
\end{equation}
where $\boldsymbol{j}^{(0)}(\boldsymbol{r})$ and $\boldsymbol{j}^{(1)}(\boldsymbol{r})$ are the unperturbed and the first-order induced current distributions, respectively. Analogous to Equation (\ref{3.2}), we may write
\begin{equation}
\boldsymbol{j}^{(0)}(\boldsymbol{r})=-ec \Psi^{(0)\dagger}(\boldsymbol{r}) \boldsymbol{\alpha} \Psi^{(0)}(\boldsymbol{r}).
\label{3.5}
\end{equation}
The expressions for the current densities in the states $\Psi(\boldsymbol{r})$ and $\Psi^{(0)}(\boldsymbol{r})$, given by \mbox{Equations (\ref{3.2})} and (\ref{3.5}), respectively, we shall rewrite as
\begin{equation}
\boldsymbol{j}(\boldsymbol{r})=-\frac{1}{2}ec \Psi^{\dagger}(\boldsymbol{r}) \boldsymbol{\alpha} \Psi(\boldsymbol{r})-\frac{1}{2}ec \Psi^{\dagger}(\boldsymbol{r}) \boldsymbol{\alpha} \Psi(\boldsymbol{r})
\label{3.6}
\end{equation}
and
\begin{equation}
\boldsymbol{j}^{(0)}(\boldsymbol{r})=-\frac{1}{2}ec \Psi^{(0)\dagger}(\boldsymbol{r}) \boldsymbol{\alpha} \Psi^{(0)}(\boldsymbol{r})-\frac{1}{2}ec \Psi^{(0)\dagger}(\boldsymbol{r}) \boldsymbol{\alpha} \Psi^{(0)}(\boldsymbol{r}).
\label{3.7}
\end{equation}
To continue the Gordon decomposition we will need Equations (\ref{2.9}) and (\ref{2.13}) rewritten in the following forms:
\begin{equation}
 \Psi(\boldsymbol{r})=\frac{\textrm{i} \hbar}{m_e c} \beta \boldsymbol{\alpha} \cdot \boldsymbol{\nabla} \Psi(\boldsymbol{r})-\frac{e}{m_e c}\beta \boldsymbol{\alpha} \cdot \boldsymbol{A}(\boldsymbol{r})\Psi(\boldsymbol{r})+\frac{1}{m_e c^2} \left[E-V_c(r)\right] \beta \Psi(\boldsymbol{r}),
\label{3.8}
\end{equation}
\begin{equation}
 \Psi^{(0)}(\boldsymbol{r})=\frac{\textrm{i} \hbar}{m_e c} \beta \boldsymbol{\alpha} \cdot \boldsymbol{\nabla} \Psi^{(0)}(\boldsymbol{r})+\frac{1}{m_e c^2} \left[E-V_c(r)\right] \beta \Psi^{(0)}(\boldsymbol{r}).
\label{3.9}
\end{equation}
Now, into the first terms on the right-hand sides of Equations (\ref{3.6}) and (\ref{3.7})  we insert the expressions (\ref{3.8}) and  (\ref{3.9}), respectively, and to the second terms we shall substitute their Hermitean conjugates. Then, carrying out some matrix algebra, we obtain 
\begin{equation}
\boldsymbol{j}(\boldsymbol{r})=
- \frac{e \hbar}{2m_e}  \boldsymbol{\nabla} \times \left[\Psi^{\dagger}(\boldsymbol{r}) \beta \boldsymbol{\Sigma} \Psi(\boldsymbol{r}) \right] 
 -\frac{e \hbar}{m_e} \Imag \left[\Psi^{\dagger}(\boldsymbol{r}) \beta \boldsymbol{\nabla} \Psi(\boldsymbol{r}) \right]
-\frac{e^2}{m_e} \Psi^{\dagger}(\boldsymbol{r}) \beta \boldsymbol{A}(\boldsymbol{r}) \Psi(\boldsymbol{r})
\label{3.10}
\end{equation}
and
\begin{equation}
\boldsymbol{j}^{(0)}(\boldsymbol{r})=
- \frac{e \hbar}{2m_e}  \boldsymbol{\nabla} \times \left[\Psi^{(0)\dagger}(\boldsymbol{r}) \beta \boldsymbol{\Sigma} \Psi^{(0)}(\boldsymbol{r}) \right]
 -\frac{e \hbar}{m_e} \Imag \left[\Psi^{(0)\dagger}(\boldsymbol{r}) \beta \boldsymbol{\nabla} \Psi^{(0)}(\boldsymbol{r}) \right].
\label{3.11}
\end{equation}
Finally, by subtracting Equation (\ref{3.11}) from Equation (\ref{3.10}) and using (\ref{2.15}), leaving only the terms at most first order, we obtain $\boldsymbol{j}^{(1)}(\boldsymbol{r})$ in the from of Equation (\ref{3.1}), with the following components:
\begin{equation}
\boldsymbol{j}_{d}^{(1)}(\boldsymbol{r})=-\frac{e^2}{m_e} \Psi^{(0)\dagger}(\boldsymbol{r}) \beta \boldsymbol{A}\Psi^{(0)}(\boldsymbol{r})
\label{3.12}
\end{equation}
and
\begin{equation}
\boldsymbol{j}_{p}^{(1)}(\boldsymbol{r})=
- \frac{e \hbar}{m_e} \boldsymbol{\nabla} \times \Real \left[\Psi^{(0)\dagger}(\boldsymbol{r}) \beta  \boldsymbol{\Sigma}\Psi^{(1)}(\boldsymbol{r}) \right]
-\frac{e \hbar}{m_e} \Imag \left[\Psi^{(0)\dagger}(\boldsymbol{r}) \beta  \boldsymbol{\nabla}\Psi^{(1)}(\boldsymbol{r})+\Psi^{(1)\dagger}(\boldsymbol{r}) \beta  \boldsymbol{\nabla}\Psi^{(0)}(\boldsymbol{r}) \right]. 
\label{3.13}
\end{equation}

\section{Dia- and paramagnetic contributions to the magnetizability}
\label{IV}
\setcounter{equation}{0}
The atomic magnetizability $\chi$ is defined through the relationship
\begin{equation}
\chi=\frac{\mu_0}{4\pi}\frac{\boldsymbol{m}^{(1)} \cdot \boldsymbol{B}}{B^2},
\label{4.1}
\end{equation}
where $\mu_0$ is the permeability of vacuum, while 
\begin{equation}
\boldsymbol{m}^{(1)}=\frac{1}{2}\int_{\mathbb{R}^3} \textrm{d}^3\boldsymbol{r} \: \boldsymbol{r} \times \boldsymbol{j}^{(1)}(\boldsymbol{r})
\label{4.2}
\end{equation}
is an induced magnetic dipole moment of the atom. If the induced current density is taken in the form of Equation (\ref{3.1}), then $\boldsymbol{m}^{(1)}$ can be represented by the sum
 \begin{equation}
\boldsymbol{m}^{(1)}=\boldsymbol{m}_{d}^{(1)}+\boldsymbol{m}_{p}^{(1)},
\label{4.3}
\end{equation}
with the components
\begin{equation}
\boldsymbol{m}_{d}^{(1)}=\frac{1}{2}\int_{\mathbb{R}^3} \textrm{d}^3\boldsymbol{r} \: \boldsymbol{r} \times \boldsymbol{j}_{d}^{(1)}(\boldsymbol{r})
\label{4.4}
\end{equation}
and
\begin{equation}
\boldsymbol{m}_{p}^{(1)}=\frac{1}{2}\int_{\mathbb{R}^3} \textrm{d}^3\boldsymbol{r} \: \boldsymbol{r} \times \boldsymbol{j}_{p}^{(1)}(\boldsymbol{r}).
\label{4.5}
\end{equation}
Consequently, the magnetizability can be divided as follows:  
\begin{equation}
\chi=\chi_{d}+\chi_{p},
\label{4.6}
\end{equation} 
where
\begin{equation}
\chi_{d}=\frac{\mu_0}{4\pi}\frac{\boldsymbol{m}_{d}^{(1)} \cdot \boldsymbol{B}}{B^2} \qquad \textrm{and} \qquad \chi_{p}=\frac{\mu_0}{4\pi}\frac{\boldsymbol{m}_{p}^{(1)} \cdot \boldsymbol{B}}{B^2}.
\label{4.7}
\end{equation}
In this section, we shall find the explicit expressions for the diamagnetic ($\chi_{d}$) and paramagnetic ($\chi_{p}$) contributions to the magnetizability of a Dirac one-electron atom being in a state characterized by the set of quantum numbers $\{ n, \kappa, \mu \}$.

To begin with the diamagnetic part, we shall insert the formulas from (\ref{2.11}) and (\ref{3.12}) into Equation (\ref{4.4}), yielding

\begin{equation}
\boldsymbol{m}_{d}^{(1)}=- \frac{e^2 B}{4m_e}\int_{\mathbb{R}^3} \textrm{d}^3\boldsymbol{r} \: \Psi^{(0)\dagger}(\boldsymbol{r}) \boldsymbol{r} \times
\left(\boldsymbol{n}_z \times \boldsymbol{r} 
\right) \beta \Psi^{(0)}(\boldsymbol{r}).
\label{4.8}
\end{equation}
Combining the above formula with the first of Equation (\ref{4.7}), we arrive at
\begin{equation}
\chi_{d}=- \frac{1}{4} \alpha^2 a_0 \int_{\mathbb{R}^3} \textrm{d}^3\boldsymbol{r} \: \Psi^{(0)\dagger}(\boldsymbol{r}) 
\left(\boldsymbol{n}_z \times \boldsymbol{r} 
\right)^2 \beta \Psi^{(0)}(\boldsymbol{r}).
\label{4.9}
\end{equation}
Putting Equation (\ref{2.4}) into Equation (\ref{4.9}), and exploiting the relation (Equation~(3.1.1) in~\cite{Szmy07}),
\begin{eqnarray}
\left( 
\boldsymbol{n}_z \cdot \boldsymbol{n}_r 
\right) 
\Omega_{\kappa \mu}(\boldsymbol{n}_r)&=&
-\frac{2\mu}{4\kappa^2-1}\Omega_{-\kappa \mu}(\boldsymbol{n}_r)
+\frac{\sqrt{\left(\kappa+\frac{1}{2}\right)^2-\mu^2}}{|2\kappa+1|}\Omega_{\kappa+1, \mu}(\boldsymbol{n}_r)
\nonumber
\\
&&+\frac{\sqrt{\left(\kappa-\frac{1}{2}\right)^2-\mu^2}}{|2\kappa-1|}\Omega_{\kappa-1, \mu}(\boldsymbol{n}_r),
\label{4.10} 
\end{eqnarray}
and after performing the angular integration, we obtain
\begin{equation}
\chi_d= - \frac{\alpha^2 a_0}{8} \: \frac{4\kappa^2+4\mu^2-1}{4\kappa^2-1} \int_0^{\infty} \textrm{d}r \: r^2 \left[P_{n\kappa}^{(0)}(r)P_{n\kappa}^{(0)}(r)-Q_{n\kappa}^{(0)}(r)Q_{n\kappa}^{(0)}(r) \right].
\label{4.11}
\end{equation}
To evaluate the radial integral we shall use Equations (\ref{2.6})--(\ref{2.7}) and exploit the recurrence formula for the Laguerre polynomials (Equation (8.971.5) in \cite{Grad94}),
\begin{equation}
L_n^{(\beta)}(\rho)=L_n^{(\beta+1)}(\rho)-L_{n-1}^{(\beta+1)}(\rho)
\label{4.12}
\end{equation}
and the following orthogonality relation (Equation (7.414.3) in \cite{Grad94}):
\begin{equation}
\int_0^{\infty} \textrm{d}\rho \: \rho^{\beta}\textrm{e}^{-\rho}L_m^{(\beta)}(\rho)L_n^{(\beta)}(\rho) 
=\frac{\Gamma(n+\beta+1)}{n!}\delta_{m n} 
\quad \quad 
[\Real \beta>-1].
\label{4.13}
\end{equation}
Also, taking into account Equation (\ref{2.2}), after some rearrangements, one obtains
\begin{equation}
\int_0^{\infty} \textrm{d}r \: r^2 \left[P_{n\kappa}^{(0)}(r)P_{n\kappa}^{(0)}(r)-Q_{n\kappa}^{(0)}(r)Q_{n\kappa}^{(0)}(r) \right]
=\frac{a_0^2}{2Z^2} (n+\gamma_{\kappa})
\left[N_{n\kappa}(5n^2+10n \gamma_{\kappa}+2\gamma_{\kappa}^2+1)-3\kappa(n+\gamma_{\kappa})
\right].
\label{4.14}
\end{equation}
Thus, the diamagnetic part of the atomic magnetizability is
\begin{equation}
\chi_d
=-\frac{\alpha^2 a_0^3}{Z^2} \: \frac{4\kappa^2+4\mu^2-1}{16\left(4\kappa^2-1\right)} \: (n+\gamma_{\kappa})
\left[N_{n\kappa}(5n^2+10n \gamma_{\kappa}+2\gamma_{\kappa}^2+1)-3\kappa(n+\gamma_{\kappa})
\right].
\label{4.15}
\end{equation}
For the ground state of the atom, i.e. when $n=0$, $\kappa=-1$, $\mu=\pm 1/2$ (and, consequently, $N_{0,-1}=1$), the above expression reduces to
\begin{equation}
\chi_d^{\textrm{(ground)}}
=-\frac{\alpha^2 a_0^3}{Z^2} \: \frac{\gamma_1 (\gamma_1+1)(2\gamma_1+1)}{12},
\label{4.16}
\end{equation}
in agreement with the corresponding formula found by Szmytkowski in ref. \cite{Szmy02b}.

Now, we will start deriving the expression for $\chi_p$ by inserting formula from (\ref{3.13}) into Equation (\ref{4.5}) and using the following relation:
\begin{equation}
 \int_{\mathbb{R}^3} \textrm{d}^3\boldsymbol{r} \: \boldsymbol{r} \times \left[\boldsymbol{\nabla} \times \boldsymbol{F}(\boldsymbol{r}) \right]=2 \int_{\mathbb{R}^3} \textrm{d}^3\boldsymbol{r} \: \boldsymbol{F}(\boldsymbol{r}) \qquad  \quad \left(r^2 \boldsymbol{F}(\boldsymbol{r}) \stackrel{r \to \infty}{\longrightarrow}0 \right).
\label{4.17}
\end{equation}
With some further algebra, we obtain the induced paramagnetic moment in the form 
\begin{equation}
\boldsymbol{m}_{p}^{(1)}=- \frac{e \hbar}{m_e} \Real \int_{\mathbb{R}^3} \textrm{d}^3\boldsymbol{r} \: \Psi^{(0)\dagger}(\boldsymbol{r}) 
\left(\boldsymbol{\hat{\Lambda}} +\boldsymbol{\Sigma} \right) \beta \Psi^{(1)}(\boldsymbol{r}).
\label{4.18}
\end{equation}
On putting the expression (\ref{2.31}) for $\Psi^{(1)}(\boldsymbol{r})$ into the above equation, then combining the resulting formula for $\boldsymbol{m}_{p}^{(1)}$ with the second of Equation (\ref{4.7}), we arrive at
\begin{equation}
\chi_p=\chi_p'+\chi_p''+\chi_p''',
\label{4.19}
\end{equation}
with the constituents
\begin{equation}
\chi_p'=  \frac{\alpha^4 a_0^3 m_e c^2 }{2}  \int_{\mathbb{R}^3} \textrm{d}^3\boldsymbol{r}  \int_{\mathbb{R}^3} \textrm{d}^3\boldsymbol{r}' \:  \Psi^{(0)\dagger}(\boldsymbol{r}) \boldsymbol{n}_z \cdot \left(\boldsymbol{\hat{\Lambda}} +\boldsymbol{\Sigma} \right) \beta \bar{G}^{(0)}(\boldsymbol{r}, \boldsymbol{r}') \beta  \boldsymbol{n}_z \cdot \left(\boldsymbol{\hat{\Lambda}'} +\boldsymbol{\Sigma} \right) \Psi^{(0)}(\boldsymbol{r}'),
\label{4.20}
\end{equation}
\begin{equation}
\chi_p''=\frac{\alpha^3 a_0^2 }{4} \Real  \int_{\mathbb{R}^3} \textrm{d}^3\boldsymbol{r}   \:  \Psi^{(0)\dagger}(\boldsymbol{r}) \boldsymbol{n}_z \cdot \left(\boldsymbol{\hat{\Lambda}} +\boldsymbol{\Sigma} \right)   \boldsymbol{n}_z \cdot \left(\boldsymbol{r} \times \boldsymbol{\alpha} \right) \Psi^{(0)}(\boldsymbol{r}),
\label{4.21}
\end{equation}
\begin{equation}
\chi_p'''=\frac{\alpha^3 a_0^2}{4}  \Real \left\{ \int_{\mathbb{R}^3} \textrm{d}^3\boldsymbol{r}   \:  \Psi^{(0)\dagger}(\boldsymbol{r}) \boldsymbol{n}_z \cdot \left(\boldsymbol{\hat{\Lambda}} +\boldsymbol{\Sigma} \right) \beta  \Psi^{(0)}(\boldsymbol{r}) \int_{\mathbb{R}^3} \textrm{d}^3\boldsymbol{r}'   \:  \Psi^{(0)\dagger}(\boldsymbol{r}') \boldsymbol{n}_z \cdot \left(\boldsymbol{r}' \times \boldsymbol{\alpha} \right) \beta \Psi^{(0)}(\boldsymbol{r}') \right\}.
\label{4.22}
\end{equation}
The expression in curly brackets on the right-hand side of Equation (\ref{4.22}) is purely imaginary because the operator $\boldsymbol{n}_z \cdot \left(\boldsymbol{\hat{\Lambda}} +\boldsymbol{\Sigma} \right)\beta$ is Hermitean, while the operator $\boldsymbol{n}_z \cdot \left(\boldsymbol{r} \times \boldsymbol{\alpha} \right) \beta$ is anti-Hermitean. Consequently, $\chi_p'''=0$, and thus
\begin{equation}
\chi_p=\chi_p'+\chi_p''.
\label{4.23}
\end{equation}

To derive the formula for $\chi_p''$, we shall substitute Equation (\ref{2.4}) into Equation (\ref{4.21}) and exploit the following relations (Equations (3.1.4), (3.1.6) and (3.2.1), respectively \cite{Szmy07}):
\begin{eqnarray} 
\left( 
\boldsymbol{n}_z \cdot \boldsymbol{\sigma} 
\right) 
\Omega_{\kappa \mu}(\boldsymbol{n}_r)=
-\frac{2\mu }{2\kappa+1}\Omega_{\kappa \mu}(\boldsymbol{n}_r)
-2\frac{\sqrt{\left(\kappa+\frac{1}{2}\right)^2-\mu^2}}{|2\kappa+1|}\Omega_{-\kappa-1, \mu}(\boldsymbol{n}_r),
\label{4.24} 
\end{eqnarray}
\begin{eqnarray}
\boldsymbol{n}_z\cdot 
\left( 
\boldsymbol{n}_r \times \boldsymbol{\sigma} 
\right) 
\Omega_{\kappa \mu}(\boldsymbol{n}_r)&=&
\textrm{i}\frac{4\mu \kappa}{4\kappa^2-1}\Omega_{-\kappa \mu}(\boldsymbol{n}_r)
+\textrm{i}\frac{\sqrt{\left(\kappa+\frac{1}{2}\right)^2-\mu^2}}{|2\kappa+1|}\Omega_{\kappa+1, \mu}(\boldsymbol{n}_r)
\nonumber
\\
&&-\textrm{i}\frac{\sqrt{\left(\kappa-\frac{1}{2}\right)^2-\mu^2}}{|2\kappa-1|}\Omega_{\kappa-1, \mu}(\boldsymbol{n}_r)
\label{4.25} 
\end{eqnarray}
and
\begin{eqnarray} 
\left( 
\boldsymbol{n}_z \cdot \hat{\boldsymbol{\Lambda}} 
\right) 
\Omega_{\kappa \mu}(\boldsymbol{n}_r)=
\frac{2\mu(\kappa+1) }{2\kappa+1}\Omega_{\kappa \mu}(\boldsymbol{n}_r)
+\frac{\sqrt{\left(\kappa+\frac{1}{2}\right)^2-\mu^2}}{|2\kappa+1|}\Omega_{-\kappa-1, \mu}(\boldsymbol{n}_r).
\label{4.26} 
\end{eqnarray}
This yields
\begin{equation}
\chi_p''=  \alpha^3 a_0^2 \: \frac{2\kappa \mu^2}{4\kappa^2-1} \int_0^{\infty} \textrm{d}r \: r P_{n\kappa}^{(0)}(r) Q_{n\kappa}^{(0)}(r). 
\label{4.27}
\end{equation}
The radial integral appearing above may be easily found with the use of Equations (\ref{2.6})--(\ref{2.7}) and (\ref{4.12})--(\ref{4.13}), with the result
\begin{equation}
\int_0^{\infty} \textrm{d}r \: r P_{n\kappa}^{(0)}(r) Q_{n\kappa}^{(0)}(r)=\frac{\alpha a_0}{4} \: \frac{2\kappa(n+\gamma_{\kappa})-N_{n\kappa}}{N_{n\kappa}}.
\label{4.28}
\end{equation}
Consequently, we arrive at
\begin{equation}
\chi_p''=\frac{\alpha^2 a_0^3}{Z^2} \: \frac{\kappa(\kappa^2-\mu^2)\mu^2}{4\kappa^2-1} \: \frac{2\kappa(n+\gamma_{\kappa})-N_{n\kappa}}{2N_{n\kappa}}.
\label{4.29}
\end{equation}

Now, let us start to evaluate the expression for the remaining component of the paramagnetic part of the magnetizability, i.e., $\chi_p'$. At first, we shall insert Equation (\ref{2.4}) and the following partial-wave expansion of the generalized Dirac--Coulomb Green function:
{\footnotesize
\begin{eqnarray}
\bar{G}\mbox{}^{(0)}(\boldsymbol{r},\boldsymbol{r}')
&=& \frac{1}{\alpha^2 a_0 m_e c^2} 
\sum_{\substack{\kappa'=-\infty \\ (\kappa'\neq0)}}^{\infty}
\sum_{\mu'=-|\kappa'|+1/2}^{|\kappa'|-1/2}\frac{1}{rr'} 
\nonumber
\\
&& \times
\left(
\begin{array}{cc}
\bar{g}\mbox{}^{(0)}_{(++)\kappa'}(r,r')
\Omega_{\kappa' \mu'}(\boldsymbol{n}_{r})
\Omega_{\kappa' \mu'}^{\dag}(\boldsymbol{n}_{r}^{\prime}) &
-\mathrm{i}\bar{g}\mbox{}^{(0)}_{(+-)\kappa'}(r,r')
\Omega_{\kappa' \mu'}(\boldsymbol{n}_{r})
\Omega_{-\kappa' \mu'}^{\dag}(\boldsymbol{n}_{r}^{\prime}) \\
\mathrm{i}\bar{g}\mbox{}^{(0)}_{(-+)\kappa'}(r,r')
\Omega_{-\kappa' \mu'}(\boldsymbol{n}_{r})
\Omega_{\kappa' \mu'}^{\dag}(\boldsymbol{n}_{r}^{\prime}) &
\bar{g}\mbox{}^{(0)}_{(--)\kappa'}(r,r')
\Omega_{-\kappa' \mu'}(\boldsymbol{n}_{r})
\Omega_{-\kappa' \mu'}^{\dag}(\boldsymbol{n}_{r}^{\prime}) 
\end{array} 
\right)
\label{4.30}
\end{eqnarray}}
into the formula given in Equation (\ref{4.20}). Then, utilizing Equations (\ref{4.24}) and (\ref{4.26}), taking into account also the orthogonality relation for the spherical spinors,
\begin{equation}
\oint_{4\pi} \textrm{d}^2\boldsymbol{n}_r \: \Omega_{\kappa \mu}^{\dagger}(\boldsymbol{n}_r) \Omega_{\kappa' \mu'}(\boldsymbol{n}_r)=\delta_{\kappa \kappa'}\delta_{\mu \mu'}
\label{4.31}
\end{equation}
to carry out the angular integrals, we obtain
\begin{equation}
\chi_p'=\chi_{p,\kappa}'+\chi_{p,-\kappa-1}'+\chi_{p,-\kappa+1}', 
\label{4.32}
\end{equation}
where we have defined
\begin{eqnarray}
\chi_{p,\kappa}'= \alpha^2 a_0^2 \cdot 2 \kappa^2 \mu^2 \int_0^{\infty}\textrm{d}r \int_0^{\infty} \textrm{d}r'  
\left( 
\begin{array}{cc} 
\frac{-1}{2\kappa+1} P_{n\kappa}^{(0)}(r)& 
\frac{1}{2\kappa-1} Q_{n\kappa}^{(0)}(r)
\end{array} 
\right)   
 \bar{\mathsf{G}}_{\kappa}^{(0)}(r,r')   
\left( 
\begin{array}{c}
\frac{-1}{2\kappa+1} P_{n\kappa}^{(0)}(r') \\*[1ex] 
\frac{1}{2\kappa-1} Q_{n\kappa}^{(0)}(r')
\end{array} 
\right),  \quad
\label{4.33}
\end{eqnarray}
\begin{eqnarray}
\chi_{p,-\kappa-1}'= \frac{\alpha^2 a_0^2}{8} \left(1-\frac{4\mu^2}{(2\kappa+1)^2} \right) \int_0^{\infty}\textrm{d}r \int_0^{\infty} \textrm{d}r'  
\left( 
\begin{array}{cc} 
 P_{n\kappa}^{(0)}(r)& 
0
\end{array} 
\right)   
 \bar{\mathsf{G}}_{-\kappa-1}^{(0)}(r,r')   
\left( 
\begin{array}{c}
 P_{n\kappa}^{(0)}(r') \\*[1ex] 
0
\end{array} 
\right)  \qquad
\label{4.34}
\end{eqnarray}
and
\begin{eqnarray}
\chi_{p,-\kappa+1}'= \frac{\alpha^2 a_0^2}{8} \left(1-\frac{4\mu^2}{(2\kappa-1)^2} \right) \int_0^{\infty}\textrm{d}r \int_0^{\infty} \textrm{d}r'  
\left( 
\begin{array}{cc} 0 &
 Q_{n\kappa}^{(0)}(r)
\end{array} 
\right)   
 \bar{\mathsf{G}}_{-\kappa+1}^{(0)}(r,r')   
\left( 
\begin{array}{c}
0 \\*[1ex] 
 Q_{n\kappa}^{(0)}(r') 
\end{array} 
\right). \qquad 
\label{4.35}
\end{eqnarray}
In the last three equations
\begin{equation}
\bar{\mathsf{G}}\mbox{}^{(0)}_{\kappa'}(r,r')
=\left(
\begin{array}{cc}
\bar{g}\mbox{}^{(0)}_{(++)\kappa'}(r,r') &
\bar{g}\mbox{}^{(0)}_{(+-)\kappa'}(r,r') \\*[1ex]
\bar{g}\mbox{}^{(0)}_{(-+)\kappa'}(r,r') &
\bar{g}\mbox{}^{(0)}_{(--)\kappa'}(r,r')
\end{array}
\right)
\label{4.36}
\end{equation}
is the generalized radial  Dirac--Coulomb Green function associated with the combined total angular momentum and parity quantum number $\kappa'$. Its Sturmian expansions, which we shall use in further calculations, are \cite{Szmy97}
\begin{equation}
\bar{\mathsf{G}}_{\kappa'}^{(0)}(r,r')
=
\sum_{n'=-\infty}^{\infty}{\frac{1}{\mu_{n'\kappa'}^{(0)}-1}}
\left( 
\begin{array}{c}
S_{n'\kappa'}^{(0)}(r) \\
T_{n'\kappa'}^{(0)}(r)
\end{array} 
\right) 
\left(
\begin{array}{cc} 
\mu_{n'\kappa'}^{(0)}S_{n'\kappa'}^{(0)}(r') & 
T_{n'\kappa'}^{(0)}(r')
\end{array}
\right) \qquad \left[\textrm{for} \quad \kappa' \neq \kappa \right]
\label{4.37}
\end{equation}
and
\begin{eqnarray}
\bar{\mathsf{G}}_{\kappa}^{(0)}(r,r')
&=&
\sum_{\substack{n'=-\infty\\(n'\neq n)}}^{\infty}{\frac{1}{\mu_{n' \kappa}^{(0)}-1}}
\left( 
\begin{array}{c}
S_{n'\kappa}^{(0)}(r) \\
T_{n'\kappa}^{(0)}(r)
\end{array} 
\right)
\left( 
\begin{array}{cc} 
\mu_{n'\kappa}^{(0)} S_{n'\kappa}^{(0)}(r') & 
T_{n'\kappa}^{(0)}(r')\end{array}\right)
\nonumber \\
&& +
\left( 
\begin{array}{c}
S_{n\kappa}^{(0)}(r) \\
T_{n\kappa}^{(0)}(r)
\end{array} 
\right)
\left( 
\begin{array}{cc} 
J_{n\kappa}^{(0)}(r') & 
K_{n\kappa}^{(0)}(r')
\end{array} 
\right)
+
\left( 
\begin{array}{c}
I_{n\kappa}^{(0)}(r) \\
K_{n\kappa}^{(0)}(r)
\end{array} 
\right)
\left( 
\begin{array}{cc} 
S_{n\kappa}^{(0)}(r') &  
T_{n\kappa}^{(0)}(r') 
\end{array} \right)
\nonumber \\
&& +
\left(
\epsilon_{n\kappa}-\frac{1}{2} 
\right)
\left( 
\begin{array}{c}
S_{n\kappa}^{(0)}(r) \\
T_{n\kappa}^{(0)}(r)
\end{array} 
\right)
\left( 
\begin{array}{cc} 
S_{n\kappa}^{(0)}(r') & 
T_{n\kappa}^{(0)}(r') 
\end{array} 
\right)
 \qquad \qquad \left[\textrm{for} \quad \kappa' = \kappa \right].
\label{4.38}
\end{eqnarray}
 In Equations (\ref{4.37})--(\ref{4.38})
\begin{eqnarray}
S_{n'\kappa'}^{(0)}(r)=\sqrt{\frac{(1+\epsilon_{n \kappa}) N_{n \kappa}(|n'|+2\gamma_{\kappa'})|n'|!}{2 Z N_{n'\kappa'}(N_{n'\kappa'}-\kappa')\Gamma(|n'|+2\gamma_{\kappa'})}}   
\left(\frac{2Zr}{a_0N_{n\kappa}}\right)^{\gamma_{\kappa'}} 
\exp\left(\frac{-Zr}{a_0N_{n\kappa}}\right)
\nonumber \\
\times 
\left[
L_{|n'|-1}^{(2\gamma_{\kappa'})}\left(\frac{2Zr}{a_0N_{n\kappa}}\right) 
+\frac{\kappa'-N_{n'\kappa'}}{|n'|+2\gamma_{\kappa'}}L_{|n'|}^{(2\gamma_{\kappa'})}\left(\frac{2Zr}{a_0N_{n\kappa}}\right)
\right]
\label{4.39}
\end{eqnarray}
and
\begin{eqnarray}
T_{n'\kappa'}^{(0)}(r)=\sqrt{\frac{(1-\epsilon_{n \kappa}) N_{n \kappa} (|n'|+2\gamma_{\kappa'}) |n'|!}{2 Z N_{n'\kappa'}(N_{n'\kappa'}-\kappa')\Gamma(|n'|+2\gamma_{\kappa'})}}   
\left(\frac{2Zr}{a_0N_{n\kappa}}\right)^{\gamma_{\kappa'}}  
\exp\left(\frac{-Zr}{a_0N_{n\kappa}}\right)
\nonumber \\
\times
\left[
L_{|n'|-1}^{(2\gamma_{\kappa'})}\left(\frac{2Zr}{a_0N_{n\kappa}}\right) 
-\frac{\kappa'-N_{n'\kappa'}}{|n'|+2\gamma_{\kappa'}}L_{|n'|}^{(2\gamma_{\kappa'})}\left(\frac{2Zr}{a_0N_{n\kappa}}\right)
\right]
\label{4.40}
\end{eqnarray}
are the radial Dirac--Coulomb Sturmian functions associated with the hydrogenic discrete state energy level, and
\begin{equation}
\mu_{n'\kappa'}^{(0)}=\frac{|n'|+\gamma_{\kappa'}+N_{n'\kappa'}}{n+\gamma_{\kappa}+N_{n \kappa}},
\label{4.41}
\end{equation}
with
\begin{equation}
N_{n'\kappa'}
=\pm\sqrt{\left(|n'|+\gamma_{\kappa'}\right)^2+(\alpha Z)^2}
=\pm \sqrt{|n'|+2|n'|\gamma_{\kappa'}+\kappa'^2}
\label{4.42}
\end{equation}
being a so-called apparent principal quantum number (it assumes the positive values if $n'>0$, and negative if $n'<0$; for $n'=0$, in the definition (\ref{4.42}), one chooses the plus sign if $\kappa'<0$, and the minus sign if $\kappa'>0$). The functions $I_{n\kappa}^{(0)}(r)$, $J_{n\kappa}^{(0)}(r)$, and $K_{n\kappa}^{(0)}(r)$, emerging in Equation (\ref{4.38}), are defined as \cite{Szmy97}
\begin{equation}
I_{n\kappa}^{(0)}(r)
=\epsilon_{n \kappa} 
\left[
-\left(\kappa+\frac{1}{2\epsilon_{n\kappa}}\right)S_{n\kappa}^{(0)}(r)
+r\left(\frac{mc(1+\epsilon_{n\kappa})}{\hbar}+\frac{\alpha Z}{r}\right)T_{n\kappa}^{(0)}(r)
\right], 
\label{4.43}
\end{equation}
\begin{equation}
J_{n\kappa}^{(0)}(r)
=\epsilon_{n \kappa} 
\left[
-\left(\kappa-\frac{1}{2\epsilon_{n\kappa}}\right)S_{n\kappa}^{(0)}(r)
+r\left(\frac{mc(1+\epsilon_{n\kappa})}{\hbar}+\frac{\alpha Z}{r} \right)T_{n\kappa}^{(0)}(r)
 \right], 
\label{4.44}
\end{equation}
\begin{equation}
K_{n\kappa}^{(0)}(r)
=\epsilon_{n \kappa} 
\left[
r\left(\frac{mc(1-\epsilon_{n\kappa})}{\hbar}-\frac{\alpha Z}{r} \right)S_{n\kappa}^{(0)}(r)
+\left(\kappa+\frac{1}{2\epsilon_{n\kappa}}\right)T_{n\kappa}^{(0)}(r) 
\right]. 
\label{4.45}
\end{equation}

If we plug the expansion from Equation (\ref{4.38}) to the right-hand side of formula in Equation (\ref{4.33}), then $\chi_{p, \kappa}'$ can be written as
\begin{equation}
\chi_{p,\kappa}'=\alpha^2 a_0^2 \: \frac{2\kappa^2\mu^2}{(4\kappa^2-1)^2} 
\left\{
R_{\kappa}^{(\infty)}+R_{\kappa}^{(a)}+R_{\kappa}^{(b)}+R_{\kappa}^{(c)} 
\right\},
\label{4.46}
\end{equation}
with the components 
\begin{eqnarray}
R_{\kappa}^{(\infty)}=\sum_{\substack{{n'=-\infty}\\(n' \neq n)}}^{\infty}{\frac{1}{\mu_{n' \kappa}^{(0)}-1}} 
\int_0^{\infty} \textrm{d}r  
\left[
(2\kappa-1)P_{n\kappa}^{(0)}(r) S_{n'\kappa}^{(0)}(r)- (2\kappa+1)Q_{n\kappa}^{(0)}(r) T_{n'\kappa}^{(0)}(r)
\right] 
\nonumber \\
\times
\int_0^{\infty} \textrm{d}r' 
\left[
(2\kappa-1) \mu_{n' \kappa}^{(0)}  P_{n\kappa}^{(0)}(r') S_{n'\kappa}^{(0)}(r')-(2\kappa+1) Q_{n\kappa}^{(0)}(r') T_{n'\kappa}^{(0)}(r')
\right],
\label{4.47}
\end{eqnarray}
\begin{eqnarray}
R_{\kappa}^{(a)}= 
\left(\epsilon_{n\kappa}-\frac{1}{2} \right) 
\int_0^{\infty} \textrm{d}r  
\left[
(2\kappa-1)P_{n\kappa}^{(0)}(r) S_{n\kappa}^{(0)}(r)-(2\kappa+1) Q_{n\kappa}^{(0)}(r) T_{n\kappa}^{(0)}(r)
\right] \nonumber \\
\times
\int_0^{\infty} \textrm{d}r'  
\left[
(2\kappa-1)P_{n\kappa}^{(0)}(r') S_{n\kappa}^{(0)}(r')-(2\kappa+1) Q_{n\kappa}^{(0)}(r') T_{n\kappa}^{(0)}(r')
\right], 
\label{4.48}
\end{eqnarray}
\begin{eqnarray}
R_{\kappa}^{(b)}=  
\int_0^{\infty} \textrm{d}r  
\left[
(2\kappa-1)P_{n\kappa}^{(0)}(r) I_{n\kappa}^{(0)}(r)-(2\kappa+1) Q_{n\kappa}^{(0)}(r) K_{n\kappa}^{(0)}(r)
\right] \nonumber \\
\times
\int_0^{\infty} \textrm{d}r'  
\left[
(2\kappa-1)P_{n\kappa}^{(0)}(r') S_{n\kappa}^{(0)}(r')-(2\kappa+1) Q_{n\kappa}^{(0)}(r') T_{n\kappa}^{(0)}(r')
\right],  
\label{4.49}
\end{eqnarray}
\begin{eqnarray}
R_{\kappa}^{(c)}=
\int_0^{\infty} \textrm{d}r  
\left[
(2\kappa-1)P_{n\kappa}^{(0)}(r) S_{n\kappa}^{(0)}(r)-(2\kappa+1) Q_{n\kappa}^{(0)}(r) T_{n\kappa}^{(0)}(r)
\right] \nonumber \\
\times
\int_0^{\infty} \textrm{d}r'  
\left[
(2\kappa-1)P_{n\kappa}^{(0)}(r') J_{n\kappa}^{(0)}(r')-(2\kappa+1) Q_{n\kappa}^{(0)}(r') K_{n\kappa}^{(0)}(r')
\right]. 
\label{4.50}
\end{eqnarray}
Exploiting Equations (\ref{4.41}) and (\ref{4.43})--(\ref{4.45}), and the following {relations:
\begin{subequations}
\begin{equation}
S_{n\kappa}^{(0)}(r)=\frac{\sqrt{a_0}N_{n\kappa}}{Z}P_{n\kappa}^{(0)}(r),
\label{4.51a}
\end{equation}
\begin{equation}
T_{n\kappa}^{(0)}(r)=\frac{\sqrt{a_0}N_{n\kappa}}{Z}Q_{n\kappa}^{(0)}(r),
\label{4.51b}
\end{equation}
\label{4.51}%
\end{subequations}
after very tedious calculations, one can show that the sum of the four components in the curly brackets in Equation (\ref{4.46}) vanishes and, consequently,
\begin{equation}
\chi_{p,\kappa}'=0.
\label{4.52}
\end{equation}

To derive the expressions for $\chi_{p,-\kappa-1}'$ and  $\chi_{p,-\kappa+1}'$, we shall insert the series expansion (\ref{4.37}) into Equations (\ref{4.34}) and (\ref{4.35}), respectively. At this stage, we obtain
\begin{equation}
\chi_{p,-\kappa+1}'= \frac{\alpha^2 a_0^2}{8} \left(1-\frac{4\mu^2}{(2\kappa-1)^2} \right) R_{-\kappa+1} 
\label{4.53}
\end{equation}
and
\begin{equation}
\chi_{p,-\kappa-1}'= \frac{\alpha^2 a_0^2}{8} \left(1-\frac{4\mu^2}{(2\kappa+1)^2} \right) R_{-\kappa-1}, 
\label{4.54}
\end{equation}
where we have defined
\begin{equation}
R_{-\kappa+1}=\sum_{n'=-\infty}^{\infty}\frac{1}{\mu_{n' \kappa'}^{(0)}-1} \left[  
\int_0^{\infty}\textrm{d}r \: Q_{n\kappa}^{(0)}(r)T_{n'\kappa'}^{(0)}(r) \right]^2  \qquad  \qquad \left[\kappa'=-\kappa+1 \right]
\label{4.55}
\end{equation}
and
\begin{equation}
R_{-\kappa-1}=\sum_{n'=-\infty}^{\infty}\frac{\mu_{n' \kappa'}}{\mu_{n' \kappa'}^{(0)}-1} \left[  
\int_0^{\infty}\textrm{d}r \: P_{n\kappa}^{(0)}(r)S_{n'\kappa'}^{(0)}(r) \right]^2  \qquad \qquad \left[\kappa'=-\kappa-1 \right].
\label{4.56}
\end{equation}

To tackle the radial integral from Equation (\ref{4.55}), we  exploit Equations (\ref{2.6}), (\ref{2.7}), (\ref{4.39}) and (\ref{4.40}) and  with the Laguerre polynomials written in the form
\begin{equation}
L_n^{(\beta)}(\rho)=\sum_{k=0}^{n} \frac{(-)^k}{k!} 
\left( 
\begin{array}{c} 
n+\beta \\
n-k 
\end{array} 
\right) 
\rho^k,
\label{4.57}
\end{equation}
and transform the integration variable according to $x=2Zr/a_0 N_{n\kappa}$, yielding
\begin{eqnarray}
\int_0^{\infty} \textrm{d}r \: 
\left[
Q_{n\kappa}^{(0)}(r) T_{n'\kappa'}^{(0)}(r)
\right]
=\xi (1-\epsilon_{n\kappa})  \Gamma(n+2\gamma_{\kappa})\sum_{k=0}^{n}\frac{(-)^k}{k!} \frac{(n-k)+(N_{n\kappa}-\kappa)}{(n-k)!\Gamma(k+2\gamma_{\kappa}+1)}
\nonumber \\
\times
\int_0^{\infty}\textrm{d}x \: x^{\gamma_{\kappa}+\gamma_{\kappa'}+k} \textrm{e}^{-x} 
\left[
L_{|n'|-1}^{(2\gamma_{\kappa'})}(x)
+\frac{N_{n'\kappa'}-\kappa'}{|n'|+2\gamma_{\kappa'}}L_{|n'|}^{(2\gamma_{\kappa'})}(x)
\right],
\label{4.58}
\end{eqnarray}
where 
\begin{equation}
\xi=\sqrt{\frac{a_0}{16 Z^2} 
\frac{n!(n+2\gamma_{\kappa})|n'|!(|n'|+2\gamma_{\kappa'}) N_{n\kappa}}
{N_{n'\kappa'} (N_{n\kappa}-\kappa)(N_{n'\kappa'}-\kappa')\Gamma(n+2\gamma_{\kappa})\Gamma(|n'|+2\gamma_{\kappa'})}}.
\label{4.59}
\end{equation}
Making use of the formula (Equation (7.414.11) in  \cite{Grad94}),
 \begin{equation}
\int_0^{\infty} \textrm{d}\rho \: \rho^{\gamma}\textrm{e}^{-\rho}L_n^{(\beta)}(\rho)
=\frac{\Gamma(\gamma+1)\Gamma(n+\beta-\gamma)}{n! \Gamma(\beta-\gamma)}
\qquad 
[\Real(\gamma)>-1]
\label{4.60}
\end{equation}
and the relation (\ref{4.42}), after some further rearrangements, we arrive at
\begin{eqnarray}
\int_0^{\infty} \textrm{d}r \: 
\left[
Q_{n\kappa}^{(0)}(r)T_{n'\kappa'}^{(0)}(r) 
\right]
=  \frac{\xi (1-\epsilon_{n\kappa}) \Gamma(n+2\gamma_{\kappa}) }{(|n'|-1)!(N_{n'\kappa'}+\kappa')} 
\sum_{k=0}^{n}\frac{(-)^k}{k!}\frac{\Gamma(|n'|+\gamma_{\kappa'}-\gamma_{\kappa}-k-1)}{(n-k)!\Gamma(k+2\gamma_{\kappa}+1)} 
\nonumber \\*[1ex]
\times
\frac{\Gamma(\gamma_{\kappa}+\gamma_{\kappa'}+k+1)}{\Gamma(\gamma_{\kappa'}-\gamma_{\kappa}-k)}  
\left[(n-k)+(N_{n\kappa}-\kappa)\right]
\left[(N_{n'\kappa'}+\kappa')+(|n'|+\gamma_{\kappa'}-\gamma_{\kappa}-k-1)
\right].
\label{4.61}
\end{eqnarray}
Inserting Equations (\ref{4.41}), (\ref{4.59}) and (\ref{4.61}) into Equation (\ref{4.55}), after some simplifications involving, among others, utilizing the trivial but useful identity,
\begin{equation}
\gamma_{\kappa'}^2-\gamma_{\kappa}^2=\kappa'^2-\kappa^2,
\label{4.62}
\end{equation}
we obtain
\begin{eqnarray}
R_{-\kappa+1}&=&\frac{a_0}{Z^2}\frac{n!\Gamma(n+2\gamma_{\kappa}+1)(n+\gamma_{\kappa}-N_{n\kappa})^2}{32 N_{n\kappa}(N_{n\kappa}-\kappa)}\sum_{k=0}^{n}\sum_{p=0}^{n}
\frac{\mathcal{Z}_{n\kappa}^{(+)}(k)\mathcal{Z}_{n\kappa}^{(+)}(p)}{\Gamma(\gamma_{\kappa'}-\gamma_{\kappa}-k)\Gamma(\gamma_{\kappa'}-\gamma_{\kappa}-p)} 
\nonumber\\*[1ex]
&& \times
\sum_{n'=\infty}^{\infty} \frac{\kappa'-N_{n' \kappa'}}{N_{n' \kappa'}} \:
\frac{\Gamma(|n'|+\gamma_{\kappa'}-\gamma_{\kappa}-k-1)\Gamma(|n'|+\gamma_{\kappa'}-\gamma_{\kappa}-p-1)}{|n'|!\Gamma(|n'|+2\gamma_{\kappa'}+1) (|n'|+\gamma_{\kappa'}-\gamma_{\kappa}-n)}  
\nonumber \\*[1ex]
&&\times  
\left[
(N_{n'\kappa'}+\kappa')+(|n'|+\gamma_{\kappa'}-\gamma_{\kappa}-k-1)
\right]  
\left[
(N_{n'\kappa'}+\kappa')+(|n'|+\gamma_{\kappa'}-\gamma_{\kappa}-p-1)\right] 
\nonumber
\\*[2ex]
 && \times (|n'|+\gamma_{\kappa'}-n-\gamma_{\kappa}-N_{n\kappa}-N_{n'\kappa'}) \qquad \qquad \qquad  \qquad \qquad \left[\kappa'=-\kappa+1 \right], 
\label{4.63}
\end{eqnarray}
where 
\begin{equation}
\mathcal{Z}_{n\kappa}^{(\pm)}(k)=\frac{(-)^{k}}{k!(n-k)!} [(n-k) \pm (N_{n\kappa}-\kappa)] \frac{\Gamma(\gamma_{\kappa}+\gamma_{1 \mp \kappa}+k+1)}{\Gamma(k+2\gamma_{\kappa}+1)}
\label{4.64}
\end{equation}
and analogously for $\mathcal{Z}_{n\kappa}^{(\pm)}(p)$.

The expression in Equation (\ref{4.63}) can be greatly simplified if, in the series $\sum_{n'=-\infty}^{\infty}(\ldots)$, one gathers together ingredients with the same absolute value of the summation index $n'$. By doing this, after much labor, with the use of Equations (\ref{4.42}) and (\ref{4.62}), one finds that
\begin{eqnarray}
R_{-\kappa+1}&=&\frac{a_0}{Z^2}\frac{n!\Gamma(n+2\gamma_{\kappa}+1)(n+\gamma_{\kappa}-N_{n\kappa})^2}{16 N_{n\kappa}(N_{n\kappa}-\kappa)}\sum_{k=0}^{n}\sum_{p=0}^{n} 
\frac{\mathcal{Z}_{n\kappa}^{(+)}(k)\mathcal{Z}_{n\kappa}^{(+)}(p)}{\Gamma(\gamma_{\kappa'}-\gamma_{\kappa}-k)\Gamma(\gamma_{\kappa'}-\gamma_{\kappa}-p)} {}
\nonumber \\
&& \times 
\sum_{n'=0}^{\infty}\frac{\Gamma(n'+\gamma_{\kappa'}-\gamma_{\kappa}-k-1)\Gamma(n'+\gamma_{\kappa'}-\gamma_{\kappa}-p-1)}{n'!\Gamma(n'+2\gamma_{\kappa'}+1)(n'+\gamma_{\kappa'}-\gamma_{\kappa}-n)}
\nonumber \\*[1ex]
&& \times 
\left\{
n'(n'+2\gamma_{\kappa'})(n'+\gamma_{\kappa'}-\gamma_{\kappa}-n)
\right.
\nonumber \\*[1ex]	
&& \quad 
-(n'+\gamma_{\kappa'}-\gamma_{\kappa}-k-1)(n'+\gamma_{\kappa'}-\gamma_{\kappa}-p-1)(n'+\gamma_{\kappa'}-\gamma_{\kappa}-n)
\nonumber \\*[1ex]	
&& \quad 
+(N_{n\kappa}-\kappa')(n'+\gamma_{\kappa'}-\gamma_{\kappa}-k-1)(n'+\gamma_{\kappa'}-\gamma_{\kappa}-p-1)
\nonumber \\*[1ex]	
&& \quad \left.
+n'(n'+2\gamma_{\kappa'}) 
\left[
(n-k-1)+(n-p-1)+(N_{n\kappa}+\kappa')
\right]	
\right\} \qquad \quad
  \left[\kappa'=-\kappa+1 \right]. \qquad
\label{4.65}
\end{eqnarray}
Further simplifications can be achieved if one notices that the sum of the two series $\sum_{n'=0}^{\infty}(\ldots )$, formed by using the first two terms in the curly braces on the right-hand side of the above equation, equals zero. Moreover, the other two series can be expressed in terms of the hypergeometric functions ${}_3F_2$ of the unit argument, with the help of the following definition \cite{Bail35,Slat66}:  
\begin{eqnarray}
 {}_3F_2 
\left(
\begin{array}{c} 
a_1, a_2, a_3\\
b_1, b_2
\end{array}
;1 
\right)
=\frac{\Gamma(b_1)\Gamma(b_2)}{\Gamma(a_1) \Gamma(a_2)\Gamma(a_3)} 
\sum_{n=0}^{\infty} \frac{\Gamma(a_1+n)\Gamma(a_2+n)\Gamma(a_3+n)}{n!\Gamma(b_1+n)\Gamma(b_2+n)}
\nonumber \\*[1ex]  
\left[\Real\left(b_1+b_2-a_1-a_2-a_3 \right)>0\right].
\label{4.66} 
\end{eqnarray}
After performing the calculations described above, Equation (\ref{4.65}) takes the form
\begin{eqnarray}
R_{-\kappa+1}&=&\frac{a_0}{Z^2}\frac{n!\Gamma(n+2\gamma_{\kappa}+1)(n+\gamma_{\kappa}-N_{n\kappa})^2}{16 N_{n\kappa}(N_{n\kappa}-\kappa)\Gamma(2\gamma_{\kappa'}+1)} 
\sum_{k=0}^{n}\sum_{p=0}^{n}\mathcal{Z}_{n\kappa}^{(+)}(k)\mathcal{Z}_{n\kappa}^{(+)}(p)
\nonumber \\*[1ex]
&& \times 
\left[
\frac{N_{n\kappa}-\kappa'}{\gamma_{\kappa'}-\gamma_{\kappa}-n} {}_3F_2 
\left(
\begin{array}{c} 
\gamma_{\kappa'}-\gamma_{\kappa}-k,\: 
\gamma_{\kappa'}-\gamma_{\kappa}-p,\: 
\gamma_{\kappa'}-\gamma_{\kappa}-n \\
\gamma_{\kappa'}-\gamma_{\kappa}-n+1,\:
2\gamma_{\kappa'}+1
\end{array}
;1 
\right)
\right.
\nonumber \\*[1ex]
&& \quad 
+
{}_3F_2 
\left(
\begin{array}{c} 
\gamma_{\kappa'}-\gamma_{\kappa}-k,\: 
\gamma_{\kappa'}-\gamma_{\kappa}-p,\: 
\gamma_{\kappa'}-\gamma_{\kappa}-n+1\\
\gamma_{\kappa'}-\gamma_{\kappa}-n+2,\:
2\gamma_{\kappa'}+1
\end{array}
;1 
\right)
\nonumber \\*[1ex]
&&  \quad \left. \times
\frac{(n-k)+(n-p)+(N_{n\kappa}+\kappa'-2)}{\gamma_{\kappa'}-\gamma_{\kappa}-n+1} 
\right] \qquad \qquad
  \left[\kappa'=-\kappa+1 \right].
\label{4.67}
\end{eqnarray}
The first of the ${}_3F_{2}$ functions appearing in Equation (\ref{4.67}) may be eliminated, if one uses the relation
\begin{eqnarray}
{}_3F_2 
\left( 
\begin{array}{c} 
a_1, a_2, a_3-1\\
a_3,b
\end{array}
;1 
\right)
=-\frac{(a_1-a_3)(a_2-a_3)}{a_3(b-a_3)}  
{}_3F_2 
\left(
\begin{array}{c} 
a_1, a_2, a_3\\
a_3+1,b
\end{array}
;1 
\right)
+\frac{\Gamma(b)\Gamma(b-a_1-a_2+1)}{(b-a_3)\Gamma(b-a_1)\Gamma(b-a_2)}
\nonumber \\ 
\left[\Real(b-a_1-a_2)>-1\right]. \quad 
\label{4.68}
\end{eqnarray}
Also, providing some further algebraic transformations, we obtain
\begin{eqnarray}
R_{-\kappa+1}&=&\frac{a_0}{Z^2}\frac{(n+\gamma_{\kappa}-N_{n\kappa})^2}{16 N_{n\kappa}(N_{n\kappa}-\kappa)(N_{n\kappa}+\kappa')}
\Bigg\{
[(n-k)+(N_{n\kappa}-\kappa)][(n-p)+(N_{n\kappa}-\kappa)]\mathcal{F}_{\kappa}^{n}(2)
\nonumber \\
&& 
+\frac{n!\Gamma(n+2\gamma_{\kappa}+1)}{(\gamma_{\kappa'}-\gamma_{\kappa}-n+1)\Gamma(2\gamma_{\kappa'}+1)}
\sum_{k=0}^n\sum_{p=0}^n \widetilde{\mathcal{Z}}_{n\kappa}^{(+)}(k)\widetilde{\mathcal{Z}}_{n\kappa}^{(+)}(p) 
\nonumber \\
&& 
\quad \times
{}_3F_2 
\left(
\begin{array}{c} 
\gamma_{\kappa'}-\gamma_{\kappa}-k,\: 
\gamma_{\kappa'}-\gamma_{\kappa}-p,\: 
\gamma_{\kappa'}-\gamma_{\kappa}-n+1 \\
\gamma_{\kappa'}-\gamma_{\kappa}-n+2,\:
2\gamma_{\kappa'}+1
\end{array}
;1 
\right)
\Bigg\} \qquad \quad
  \left[\kappa'=-\kappa+1 \right], \qquad \quad
\label{4.69}
\end{eqnarray}
where we have defined
\begin{equation}
\widetilde{\mathcal{Z}}_{n\kappa}^{(\pm)}(k)=
\left
[(n-k) \pm (N_{n \kappa}-\kappa)\right]
\mathcal{Z}_{n\kappa}^{(\pm)}(k)
\label{4.70}
\end{equation}
and similarly for $\widetilde{\mathcal{Z}}_{n\kappa}^{(\pm)}(p)$. The function $\mathcal{F}_{\kappa}^{n}(M)$, also appearing in Equation (\ref{4.69}), was defined by us in the Appendix to ref. \cite{Stef15} in Equation (A.1). Basing on the analysis carried out in that Appendix, with some algebra, one can show that
\begin{eqnarray}
[(n-k)+(N_{n\kappa}-\kappa)][(n-p)+(N_{n\kappa}-\kappa)]\mathcal{F}_{\kappa}^{n}(2)=2(N_{n\kappa}-\kappa)\left[n(n+2\gamma_{\kappa})+\kappa-2(n+\gamma_{\kappa})N_{n\kappa}\right]. \quad
\label{4.71}
\end{eqnarray}

The formula for $R_{-\kappa-1}$ may be derived in the similar way. Generally, we found that
\begin{eqnarray}
R_{-\kappa\pm1}&=&\frac{a_0}{Z^2} \frac{(n+\gamma_{\kappa}\mp N_{n \kappa})^2}{128N_{n \kappa}(N_{n \kappa}+\kappa')}
\Big[ 
2\kappa-4(n+\gamma_{\kappa})N_{n\kappa}\pm 2n(n+2\gamma_{\kappa})
\nonumber \\
&& +\frac{n!\Gamma(n+2\gamma_{\kappa}+1)}{(N_{n\kappa}-\kappa)(\gamma_{\kappa'}-\gamma_{\kappa}-n+1)\Gamma(2\gamma_{\kappa'}+1)} 
\sum_{k=0}^n\sum_{p=0}^n\widetilde{\mathcal{Z}}_{n\kappa}^{(\pm)}(k)\widetilde{\mathcal{Z}}_{n\kappa}^{(\pm)}(p)
\nonumber \\
&& \quad 
\left.
\times {}_3F_2 
\left(
\begin{array}{c} 
\gamma_{\kappa'}-\gamma_{\kappa}-k,\: 
\gamma_{\kappa'}-\gamma_{\kappa}-p,\: 
\gamma_{\kappa'}-\gamma_{\kappa}-n+1\\
\gamma_{\kappa'}-\gamma_{\kappa}-n+2,\:
2\gamma_{\kappa'}+1
\end{array}
;1 
\right) 
\right] \qquad \quad [\kappa'=-\kappa \pm 1]. \quad \quad
\label{4.72}
\end{eqnarray}
Inserting the last result  to Equations (\ref{4.53}) and (\ref{4.54}), we arrive at
\begin{eqnarray}
\chi_{p,-\kappa \pm 1}'&=&\frac{\alpha^2 a_0^3}{Z^2} \left(1-\frac{4\mu^2}{(2\kappa \mp 1)^2} \right) \frac{(n+\gamma_{\kappa} \mp N_{n \kappa})^2}{128N_{n \kappa}(N_{n \kappa}+\kappa')}
\Bigg[ 
2\kappa-4(n+\gamma_{\kappa})N_{n\kappa}\pm 2n(n+2\gamma_{\kappa})
\nonumber \\
&& +\frac{n!\Gamma(n+2\gamma_{\kappa}+1)}{(N_{n\kappa}-\kappa)(\gamma_{\kappa'}-\gamma_{\kappa}-n+1)\Gamma(2\gamma_{\kappa'}+1)} 
\sum_{k=0}^n\sum_{p=0}^n\widetilde{\mathcal{Z}}_{n\kappa}^{(\pm)}(k)\widetilde{\mathcal{Z}}_{n\kappa }^{(\pm)}(p)
\nonumber \\
&&\quad
\times {}_3F_2 
\left(
\begin{array}{c} 
\gamma_{\kappa'}-\gamma_{\kappa}-k,\: 
\gamma_{\kappa'}-\gamma_{\kappa}-p,\: 
\gamma_{\kappa'}-\gamma_{\kappa}-n+1\\
\gamma_{\kappa'}-\gamma_{\kappa}-n+2,\:
2\gamma_{\kappa'}+1
\end{array}
;1 
\right) 
\Bigg] \qquad \quad [\kappa'=-\kappa \pm 1]. \qquad \quad
\label{4.73}
\end{eqnarray}
Splitting together Equations (\ref{4.32})  and (\ref{4.52}), one may write
\begin{equation}
\chi_{p}'=\chi_{p,-\kappa+1}'+\chi_{p,-\kappa-1}',
\label{4.74}
\end{equation}
where both constituents appearing on the right-hand side of the above equation are defined by Equation\  (\ref{4.73}).
Finally, combining Equations (\ref{4.23}), (\ref{4.29}), (\ref{4.73}) and (\ref{4.74}), we found the paramagnetic contribution to the atomic magnetizability in the following explicit form:
\begin{eqnarray}
\chi_p &=& \frac{\alpha^2 a_0^3}{Z^2} \:  \Bigg\{\frac{\kappa(\kappa^2-\mu^2)\mu^2}{4\kappa^2-1} \: \frac{2\kappa(n+\gamma_{\kappa})-N_{n\kappa}}{2N_{n\kappa}}
 + \sum_{+,-} \left(1-\frac{4\mu^2}{(2\kappa \mp 1)^2} \right) \frac{(n+\gamma_{\kappa} \mp N_{n \kappa})^2}{128N_{n \kappa}(N_{n \kappa}-\kappa \pm 1)}
\nonumber \\
&& 
\times
\Bigg[ 
2\kappa-4(n+\gamma_{\kappa})N_{n\kappa}\pm 2n(n+2\gamma_{\kappa})
+\frac{n!\Gamma(n+2\gamma_{\kappa}+1)}{(N_{n\kappa}-\kappa)(\gamma_{1 \mp \kappa}-\gamma_{\kappa}-n+1)\Gamma(2\gamma_{1 \mp \kappa}+1)}  \nonumber \\
&&\quad
\times
\sum_{k=0}^n\sum_{p=0}^n\widetilde{\mathcal{Z}}_{n\kappa }^{(\pm)}(k)\widetilde{\mathcal{Z}}_{n\kappa }^{(\pm)}(p) \:
 {}_3F_2 
\left(
\begin{array}{c} 
\gamma_{1 \mp \kappa}-\gamma_{\kappa}-k,\: 
\gamma_{1 \mp \kappa}-\gamma_{\kappa}-p,\: 
\gamma_{1 \mp \kappa}-\gamma_{\kappa}-n+1\\
\gamma_{1 \mp \kappa}-\gamma_{\kappa}-n+2,\:
2\gamma_{1 \mp \kappa}+1
\end{array}
;1 
\right) 
\Bigg]
\Bigg\},
\nonumber
\\
\label{4.75}
\end{eqnarray}
where $\sum_{+,-}(\ldots)$ denotes  that the components with the upper and the lower sign should be added together. On putting $n=0$, $\kappa=-1$ and $\mu= \pm  1/2$ into Equation (\ref{4.75}), then transforming the resulting $ {}_3F_2 $ function with the help of the following relation:
\begin{eqnarray}
{}_3F_2 
\left( 
\begin{array}{c} 
a_1+1, a_2+1, a_3+1\\
a_3+2,b
\end{array}
;1 
\right)
=\frac{a_3+1}{a_1 a_2}\Bigg[(a_3-b+1)
{}_3F_2 
\left(
\begin{array}{c} 
a_1, a_2, a_3\\
a_3+1,b
\end{array}
;1 
\right)
\nonumber
\\
+\frac{\Gamma(b)\Gamma(b-a_1-a_2-1)}{\Gamma(b-a_1)\Gamma(b-a_2)} \big[(b-a_1-1)(b-a_2-1)-a_3 (b-a_1-a_2-1)\big] \Bigg]
\nonumber \\ 
\left[\Real(b-a_1-a_2)>-1\right] 
\label{4.76}
\end{eqnarray}
and some further algebraic transformations, we obtain the expression for  $\chi_p$ of the atom in the ground state in the form 
\begin{eqnarray}
\chi_p^{\textrm{(ground)}} &=& -\frac{\alpha^2 a_0^3}{Z^2} \Bigg[ \frac{(\gamma_1+1)(2\gamma_1+1)(\gamma_1-2)}{36}+\frac{\Gamma^2(\gamma_1+\gamma_2+2)}{72(\gamma_2-\gamma_1)\Gamma(2\gamma_1+1)\Gamma(2\gamma_2+1)} {}
\nonumber
\\
&& 
\times
{}_3F_2 
\left( 
\begin{array}{c} 
\gamma_2-\gamma_1-1, \gamma_2-\gamma_1-1, \gamma_2-\gamma_1\\
\gamma_2-\gamma_1+1,2\gamma_2+1
\end{array}
;1 
\right)
\Bigg]. 
\label{4.77}
\end{eqnarray}
The above result is identical to the corresponding formula derived in ref. \cite{Szmy02b}. Moreover, by using Equations (\ref{4.15}) and (\ref{4.75}), one may verify that, as might be expected, the sum $\chi_d+\chi_p$ coincides with the total magnetizability of the relativistic one-electron atom in an arbitrary discrete energy state, found by us some time ago \cite{Stef15,Stef16}.
 
Based on Formulas (\ref{4.6}), (\ref{4.15}), (\ref{4.29}) and (\ref{4.73})--(\ref{4.75}), we have also performed numerical calculations showing relative dia- and paramagnetic contributions to magnetizabilities for Dirac one-electron atoms with a wide range of nuclear charges, $1 \leqslant Z \leqslant 137$. The appropriate results have been already published by us in ref. \cite{Stef20}.

\section{Conclusions}
\label{V}
\setcounter{equation}{0}

In this work, we derived analytically closed-form expressions for the diamagnetic ($\chi_{d}$) and paramagnetic ($\chi_{p}$) contributions to the magnetizability of the relativistic one-electron atom in an arbitrary discrete energy state. We carried out the Gordon decomposition approach combined with the perturbation theory. In calculations, we have used the Sturmian expansion of the generalized Dirac--Coulomb Green function \cite{Szmy97}.   Our general result for $\chi_{p}$ has the form of a double finite sum involving the generalized hypergeometric functions ${}_3F_2$ of the unit argument; for the ground state it reduces to the formula found by Szmytkowski some time ago \cite{Szmy02b}. The expression for $\chi_{d}$ is of an elementary form, and also agrees with the corresponding result provided in the aforementioned article. 

This paper may be treated as a prequel to our recent work \cite{Stef20}, in which the results of numerical calculations of $\chi_{d}$ and $\chi_{p}$  for selected hydrogenic ions in some atomic states are tabulated. It should be mentioned here that, for the atomic ground state, our numerical results nearly perfectly agree with the appropriate values given in \cite{Szmy02b}, while for the excited states in the above mentioned article we have drawn some interesting conclusions about the impact of relativity on the obtained values.

This work provides also further evidence of the usefulness of the Sturmian expansion of the generalized Dirac--Coulomb Green function \cite{Szmy97} in the relativistic calculations of electromagnetic properties of the relativistic hydrogenlike ions.


\begin{thebibliography}{99}
\bibitem{Saku67}
		J.\ J.\ Sakurai
		1967
		{\it Advanced Quantum Mechanics}
		(Massachusetts: Addison-Wesley, Reading)
\bibitem{Gord28}
		W.\ Gordon
		1928
		Der Strom der Diracschen Elektronentheorie 
		{\it Z.\ Phys.} {\bf 50} 630--2 
\bibitem{Vleck32}
		J.\ H.\ Van Vleck 
		1932
		{\it The theory of electric and magnetic susceptibilities} 
		(London: Oxford University Press) 
\bibitem{Hodg14}
		W.\ B.\ Hodge, S.\ V.\ Migirditch and W.\ C.\ Kerr
		2014
		Electron spin and probability current density in quantum mechanics
		{\it Am.\ J. Phys.} {\bf 82} 681--90
\bibitem{Summ24}
		F.\ F.\ Summa
		2024
		{\it Molecular properties via induced current densities}
		Springer Theses, Springer
\bibitem{Ster62}
		M.\ M.\  Sternheim
		1962
		Second-Order Effects of Nuclear Magnetic Fields
		{\it Phys.\ Rev.} {\bf 128} 676--7 
\bibitem{Pyyk77}
		P. Pyykk{\"o}
		1977
		Relativistic theory of nuclear spin-spin coupling in molecules
		{\it Chem.\ Phys.} {\bf 22} 289--96 
\bibitem{Pyyk83}
		P. Pyykk{\"o}
		1983
		On the relativistic theory of NMR chemical shifts
		{\it Chem.\ Phys.} {\bf 74} 1--7 
\bibitem{Auca99}
		G.\ A.\ Aucar, T.\ Saue, L.\ Visscher and H.\ A.\ Jensen
		1999
		On the origin and contribution of the diamagnetic term in four-component relativistic calculations of magnetic properties
		{\it J.\ Chem.\ Phys.} {\bf 110} 6208--18 
\bibitem{Kutz03}
		W.\ Kutzelnigg
		2003
		Diamagnetism in relativistic theory
		{\it Phys.\ Rev.\ A} {\bf 67} 032109 
\bibitem{Szmy02}
		R.\ Szmytkowski
		2002
		Larmor diamagnetism and Van Vleck paramagnetism in relativistic quantum theory: The Gordon decomposition approach
		{\it Phys.\ Rev.\ A} {\bf 65} 032112 
\bibitem{Pype83}
		N.\ C.\ Pyper
		1983
		The relativistic theory of the chemical shift
		{\it Chem.\ Phys.\ Lett.} {\bf 96} 204--10 
\bibitem{Pype83a}
		N.\ C.\ Pyper
		1983
		The Breit interaction in external magnetic fields
		{\it Chem.\ Phys.\ Lett.} {\bf 96} 211--17
\bibitem{Pype88}
		N.\ C.\ Pyper
		1988
		Exact relativistic analogues of the non-relativistic hyperfine structure operators
		{\it Mol.\ Phys.} {\bf 64} 933--61
\bibitem{Pype99}
		N.\ C.\ Pyper
		1999
		Relativistic theory of nuclear shielding in one-electron atoms: 1. Theoretical foundations and first-order terms
		{\it Mol.\ Phys.} {\bf 97} 381--90
\bibitem{Pype99a}
		1999
		N.\ C.\ Pyper and Z.\ C.\ Zhang
		Relativistic theory of nuclear shielding in one-electron atoms: 2. Analytical and numerical results
		{\it Mol.\ Phys.} {\bf 97} 391--413
\bibitem{Batt01}
		M.\ Battocletti and H.\ Ebert
		2001
		Decomposition of the relativistic hyperfine interaction operator: Application to the ferromagnetic alloy systems fcc $\textrm{Fe}_{x}\textrm{Ni}_{1-x}$, fcc $\textrm{Fe}_{x}\textrm{Pd}_{1-x}$, and fcc $\textrm{Co}_{x}\textrm{Pt}_{1-x}$
		{\it Phys.\ Rev.\ B} {\bf 64} 094417	
\bibitem{Szmy97}
		R.\ Szmytkowski
		1997
		The Dirac--Coulomb Sturmians and the series expansion of the Dirac--Coulomb Green function: application to the relativistic polarizability of the hydrogen-like atom
		{\it J.\ Phys.\ B} {\bf 30} 825--61 
\bibitem{Stef15}
		P.\ Stefa{\'n}ska
		2015
		Magnetizability of the relativistic hydrogenlike atom in an arbitrary discrete energy eigenstate: Application of the Sturmian expansion of the generalized Dirac--Coulomb Green function 
		{\it Phys.\ Rev.\ A} {\bf 92} 032504  
\bibitem{Stef16}
		P.\ Stefa{\'n}ska
		2016
		Magnetizabilities of relativistic hydrogenlike atoms in some arbitrary discrete energy eigenstates
		{\it At.\ Data Nucl.\ Data Tables} {\bf 108} 193--210 
\bibitem{Stef16a}
		P.\ Stefa{\'n}ska
		2016
		Magnetic-field-induced electric quadrupole moments for relativistic hydrogenlike atoms: Application of the Sturmian expansion of the generalized Dirac–-Coulomb Green function 
		{\it Phys.\ Rev.\ A} {\bf 93} 022504 
\bibitem{Stef16b}
		P.\ Stefa{\'n}ska
		2016
		Closed-form expression for the magnetic shielding constant of the relativistic hydrogenlike atom in an arbitrary discrete energy eigenstate: Application of the Sturmian expansion of the generalized Dirac--Coulomb Green function 
		{\it Phys.\ Rev.\ A} {\bf 94} 012508 
\bibitem{Stef17}
		P.\ Stefa{\'n}ska
		2017
		Magnetic-dipole-to-electric-quadrupole cross-susceptibilities for relativistic hydrogenlike atoms in some low-lying discrete energy eigenstates
		{\it At.\ Data Nucl.\ Data Tables} {\bf 113} 316--49  
\bibitem{Stef18}
		P.\ Stefa{\'n}ska
		2018
		Nuclear magnetic shielding constants of Dirac one-electron atoms in some low-lying discrete energy eigenstates
		{\it At.\ Data Nucl.\ Data Tables} {\bf 120} 352--72 
\bibitem{Szmy02b}
		R.\ Szmytkowski
		2002
		Magnetizability of the relativistic hydrogen-like atom: application of the Sturmian expansion of the first-order Dirac--Coulomb Green function
		{\it J.\ Phys.\ B} {\bf 35} 1379--91 
\bibitem{Szmy04}
		R.\ Szmytkowski and K.\ Mielewczyk
		2004
		Gordon decomposition of the static dipole polarizability of the relativistic hydrogen-like atom: application of the Sturmian expansion of the first-order Dirac--Coulomb Green function
		{\it J.\ Phys.\ B} {\bf 37} 3961--72
\bibitem{Szmy11}
		R.\ Szmytkowski and P.\ Stefa{\'n}ska
		2011
		Comment on ``Four-component relativistic theory for NMR parameters: Unified formulation and numerical assessment of different approaches''
		{\it e-print arXiv:1102.1811}
\bibitem{Stef12}
		P.\ Stefa{\'n}ska and R.\ Szmytkowski
		2012
		Electric and magnetic dipole shielding constants for the ground state of the relativistic hydrogen-like atom: Application of the Sturmian expansion of the generalized Dirac--Coulomb Green 	function
		{\it Int.\ J.\ Quantum Chem.} {\bf 112} 1363--72 
\bibitem{Szmy12}
		R.\ Szmytkowski and P.\ Stefa{\'n}ska
		2012
		Magnetic-field-induced electric quadrupole moment in the ground state of the relativistic hydrogenlike atom: Application of the Sturmian expansion of the generalized Dirac--Coulomb Green function 
		{\it Phys.\ Rev.\ A} {\bf 85} 042502  
\bibitem{Szmy14}
    R.\ Szmytkowski and P.\ Stefa{\'n}ska
		2014
    Electric-field-induced magnetic quadrupole moment in the ground state of the relativistic hydrogenlike atom: Application of the Sturmian expansion of the generalized Dirac--Coulomb Green function 
		{\it Phys.\ Rev.\ A} {\bf 89} 042502

\bibitem{Szmy07}
		R.\ Szmytkowski
		2007
		Recurrence and differential relations for spherical spinors
		{\it J.\ Math.\ Chem.} {\bf 42} 397--413
\bibitem{Magn66}
		W.\ Magnus, F.\ Oberhettinger, and R.\ P.\ Soni
		1966
		{\it Formulas and Theorems for the Special Functions of Mathematical Physics, 3rd ed.} 
		(Berlin: Springer)
\bibitem{Schi55}
		See, for example,
		L.\ I.\ Schiff,
		1955
		{\it Quantum Mechanics, 2nd ed.}
		(London: McGraw-Hill) 
\bibitem{Grad94}
		I.\ S.\ Gradshteyn and I.\ M.\ Ryzhik
		1994
		{\it Table of Integrals, Series, and Products, 5th ed.}
		(San Diego: Academic) 				
\bibitem{Bail35}
		W.\ N.\ Bailey
		1935
		{\it Generalized Hypergeometric Series}
		(Cambridge: Cambridge University Press)
		[reprint: 1972 New York: Hafner]
\bibitem{Slat66}
		L.\ J.\ Slater
		1966
		{\it Generalized Hypergeometric Functions}
		(Cambridge: Cambridge University Press)
\bibitem{Stef20}
		P.\ Stefa{\'n}ska
		2020
		Dia- and paramagnetic contributions to magnetizabilities of relativistic hydrogenlike atoms in some low-lying discrete energy eigenstates
		{\it At.\ Data Nucl.\ Data Tables} {\bf 135--136} 101360  
\end{thebibliography}
\end{document}